\documentclass[twocolumn]{aastex63}
\usepackage{amsmath}
\usepackage{graphicx}
\usepackage{booktabs}
\usepackage{bm}

\newcommand{\mbh}{M_{\rm BH}}
\newcommand{\lbol}{L_{\rm bol}}
\newcommand{\alplam}{\alpha_{\lambda}}
\newcommand{\ebv}{E(B-V)}
\newcommand{\redd}{R_{\rm Edd}}

\shorttitle{Red Type-1 Quasars after Cosmic Noon}
\shortauthors{Y. Kim et al.}

\begin{document}

\title{Red Type-1 Quasars after Cosmic Noon and Impact on $L_{\rm UV}$-related Quasar Statistics}

\correspondingauthor{Dohyeong Kim}
\email{yjkim.ast@gmail.com; dh.dr2kim@gmail.com}

\author[0000-0003-1647-3286]{Yongjung Kim}
\affiliation{Korea Astronomy and Space Science Institute, Daejeon 34055, Republic of Korea}
\affiliation{Department of Astronomy and Atmospheric Sciences, College of Natural Sciences, Kyungpook National University, Daegu 41566, Republic of Korea}

\author[0000-0002-6925-4821]{Dohyeong Kim}
\affiliation{Department of Earth Sciences, Pusan National University, Busan 46241, Republic of Korea}

\author[0000-0002-8537-6714]{Myungshin Im}
\affiliation{SNU Astronomy Research Center, Seoul National University, 1 Gwanak-ro, Gwanak-gu, Seoul 08826, Republic of Korea}
\affiliation{Astronomy Program, Department of Physics \& Astronomy, Seoul National University, 1 Gwanak-ro, Gwanak-gu, Seoul 08826, Republic of Korea}

\author[0000-0002-3560-0781]{Minjin Kim}
\affiliation{Department of Astronomy and Atmospheric Sciences, College of Natural Sciences, Kyungpook National University, Daegu 41566, Republic of Korea}





\begin{abstract}

Over the past decades, nearly a million quasars have been explored to shed light on the evolution of supermassive black holes and galaxies.
The ultraviolet-to-optical spectra of type-1 quasars particularly offer insights into their black hole activities. 
Recent findings, however, raise questions about the prevalence of red type-1 quasars of which colors might be due to dust-obscuration and their potential influence on luminosity-related properties of quasars. 
We examine the fraction of red type-1 quasars within the redshift range of $0.68\leq z < 2.20$, applying spectral energy distribution (SED) fitting using optical-to-MIR photometric data of Sloan Digital Sky Survey Data Release 14 quasars. 
Approximately 10\,\% of the type-1 quasars exhibit red colors suggestive of dust obscuration.
There is an association between the brightness of the MIR luminosity and a higher fraction of red type-1 quasars, albeit with negligible redshift evolution.
By employing $\ebv$ values from the SED fitting, we obtained dereddened luminosity of the red type-1 quasars and reassess the quasar luminosity function (QLF) and black hole mass ($\mbh$) estimates.
Result shows a modest increase in the number density of bright quasars, linking to more flatten bright-end slope of QLFs, while $\mbh$ adjustments are minimal.
Current SDSS selections in optical could miss a significant population of heavily dust-obscured quasars.
As future MIR surveys like SPHEREx expand, they may reveal enough obscured quasars to prompt a more profound revision of fundamental quasar properties.

\end{abstract}



\section{Introduction \label{sec:intro}}

In the past two decades, significant advancements in observational techniques and the rise of large-scale surveys have propelled quasar research into a golden age. 
Notably, the Sloan Digital Sky Survey (SDSS; \citealt{York00,Gunn06}) has played a pivotal role in this endeavor, extensively cataloging various type-1 quasars across cosmic epochs \citep{Schneider02,Schneider03,Schneider05,Schneider07,Schneider10,Paris12,Paris14,Paris17,Paris18,Lyke20}. 
These catalogs have undergone continuous updates, with the latest iteration by \cite{Lyke20} based on SDSS Data Release 16 (DR16) containing over 0.7 million quasars spanning a sky area of 14,000 deg$^{2}$. 
These vast datasets have empowered astronomers to statistically investigate the nature of quasars with unprecedented precision.

In particular, such a large number of quasars enables us to refine our comprehension of quasar demographics, described by the quasar luminosity function (QLF; \citealt{Richards06a, Jiang08,Jiang09,Jiang16, McGreer13, Palanque13,Palanque16, Ross13}), across cosmic time.
Unlike galaxy populations described by a Schechter function (e.g., \citealt{Blanton01,Montero09}), it  has been revealed that the QLF is canonically well described as a broken power-law function over wide redshift ranges \citep{Kulkarni19,Shen20,KimIm21}.
These comprehensive studies suggest that the quasar number density, which exhibits density evolution according to dark matter halo mass evolution from cosmic dawn ($z\gtrsim6$), peaks at $z\sim2$, a period often referred to as cosmic noon, and then shows a luminosity evolution resulting in a decrease in the number of bright quasars.

On the other hand, the large and homogenous spectral datasets of type-1 quasars allow statistical investigation of their central supermassive black hole (SMBH) activities \citep{Shen11, Rakshit20}.
Under the assumption of virial motion of line-emitting gas around the black hole (BH), spectral features of emission line components provide dynamics around the central engine and the resultant BH activities.
Specifically, an enormous amount of materials are accreting into the central BH in the sub-Eddington levels, while non-negligible fraction of them are radiating in their maximum level reaching (or over) the Eddington limit.

Amid the wealth of discoveries, new inquiries have surfaced, challenging our understanding of quasar behavior. 
Notably, recent findings have brought into question the prevalence of red type-1 quasars in optical surveys, whose distinctive colors may be attributed to dust-obscuration within their host galaxies. 
\cite{Kim23} reported that 16\,\% of SDSS type-1 quasars at $z<0.5$ are likely obscured.
This inference was drawn from the discrepancy between the bolometric luminosities derived from the optical and mid-infrared (MIR) luminosities (MIR luminosity is relatively immune from the effects of dust extinction, highlighting its significance in such analyses).
This finding contrasts with earlier beliefs that optically selected type-1 quasars were minimally affected by dust obscuration, unlike the high fraction of obscured quasars in X-ray and infrared active galactic nuclei (AGN) surveys (e.g., \citealt{Polletta08,Lacy13,Lacy15,Ueda14,Aird15,Glikman18}).

Under the merger-driven quasar evolutionary scenario, dust-obscured quasars are in the actively accreting phase, known as the blowout phase, which precedes normal quasars without obscuration in optical (e.g., \citealt{Hopkins08}).
In this paradigm, red type-1 quasars in optical surveys could serve as promising sources for the preceding population, akin to the red quasars selected in infrared surveys \citep{Urrutia09,Urrutia12,Glikman12,Glikman18,KimD15,KimIm18}.
To verify whether the red type-1 quasars in optical surveys are in the blowout phase, various types of evidence are required (\citealt{Urrutia08,Urrutia09,Urrutia12,Glikman15,KimD15,Kim18,KimIm18,Kim24a}): high Eddington ratios ($\redd$), dust-originated red colors, high star-formation rates, and potential merging features.
However, primarily, it is necessary to ascertain whether they exhibit high accretion rates, with careful consideration of potential underestimation of the rate due to obscuration.

\cite{Banerji15} suggests that the number density of extremely obscured quasars with $0.5<\ebv<1.5$ at $z>2$ are underestimated due to high obscuration.
When considering the dereddening effect, their QLF appears to flatten more at the bright end compared to normal quasars, implying a high number of actively accreting BHs.
Additionally, the BH mass ($\mbh$) function and $\redd$ distributions conducted at $z<2$ \citep{Kelly13} indicate an earlier accreting phase and possibly obscured growth of the BHs.
Inspired by the non-negligible fraction of red type-1 quasars even at $z<0.5$ \citep{Kim23}, as well as the existence of extremely obscured quasars at $z>2$ \citep{Banerji15}, it is evident that red type-1 quasars also persist after the cosmic noon ($z\lesssim2$).
This persistence may offer insight into the composition of quasar population, prompting further investigation into their potential influence on the luminosity-related quasar statistics.

In this paper, we study the presence of red type-1 quasars and their implications to the luminosity-related quasar properties.
We describe our data and analysis including SED fitting in Section \ref{sec:sample}.
The classification of red type-1 quasars and their fraction within the type-1 quasar sample are presented in Section \ref{sec:redtype1}.
Subsequently, we examine how the prevalence of the red type-1 quasars influences the luminosity-related statistics of type-1 quasars in Section \ref{sec:impact}.
We discuss possible issues with our method in Section \ref{sec:discussion}.
Through this paper, we assume the standard $\Lambda$CDM universe with cosmological coefficients of $H_{0}=70$ km s$^{-1}$ Mpc$^{-1}$, $\Omega_{m}=0.3$ and $\Omega_{\Lambda}=0.7$.
All the magnitudes are given in AB system.

\section{Data and Analysis \label{sec:sample}}

\subsection{SDSS DR14Q \label{sec:dr14q}}

We use the catalog of SDSS DR14 quasars (hereafter DR14Q) as presented in \cite{Paris18}, consisting of 526,356 quasars over 9,376\,deg$^{2}$ area\footnote{In this study, we utilize the catalog of DR14Q instead of that from the SDSS DR16 data \citep{Lyke20}, as the latter does not include the W3 and W4 band photometry data, despite being the latest release. Note that the difference between the two samples following our selection criteria is negligible.}.
All the DR14Q were identified as type-1 quasars through spectroscopy and they span the redshift range of $0<z<7$.
The catalog provides not only the photometric/spectroscopic properties obtained from SDSS data itself, but also those from the positionally matched survey data covering various wavelengths from X-ray to radio.
For example, most of them are detected by the Wide-field Infrared Survey Explorer (WISE; \citealt{Wright10}) and their W1 to W4 magnitudes from the AllWISE data release Point Source Catalog\footnote{\url{https://wise2.ipac.caltech.edu/docs/release/allwise/}} (AllWISE PSC; \citealt{Wright10,Mainzer11,Cutri21}) are also given.

For the spectra of the DR14Q, \cite{Rakshit20} performed multi-component spectral fitting using \texttt{PyQSOFit}\footnote{\url{https://github.com/legolason/PyQSOFit}}, a Python package for fitting quasar spectrum \citep{Guo18}.
They provide the best-fit parameters for the continuum and line emission properties of each quasar.
For instance, they provide the power-law continuum slope of $\alpha_{\lambda}$, defined as 

\begin{equation}
f_{\rm PL}(\lambda) = f_{\rm PL,0}(\lambda/\lambda_{0})^{\alpha_{\lambda}}, \label{equ:fpl}
\end{equation}

\noindent where $f_{\rm PL}(\lambda)$ is the power-law continuum at the reference wavelength of $\lambda_{0}= 3000\,\rm\AA{}$ with the normalization factor of $f_{\rm PL,0}$.
They modeled the entire continuum by combining this $f_{\rm PL}(\lambda)$ with the Balmer continuum (e.g., \citealt{Dietrich02}) and \ion{Fe}{2} multiplet (e.g., \citealt{Vestergaard01}) components.
The monochromatic luminosity at a given wavelength in units of erg s$^{-1}$ is then obtained from this entire continuum model ($\lambda L_{\lambda}$ at 3000\,$\rm \AA{}$, hereafter shortly $L_{\rm 3000}$).
Moreover, they measured the spectral properties (e.g., equivalent width (EW), full width at half maximum (FWHM), line luminosity) of various emission lines (e.g., \ion{C}{4}, \ion{Mg}{2}, H$\beta$, H$\alpha$, etc.).
Note that \cite{Rakshit20} subtracted the host galaxy light before the spectral fitting.

\begin{figure*}
\centering
\epsscale{1.1}
\plotone{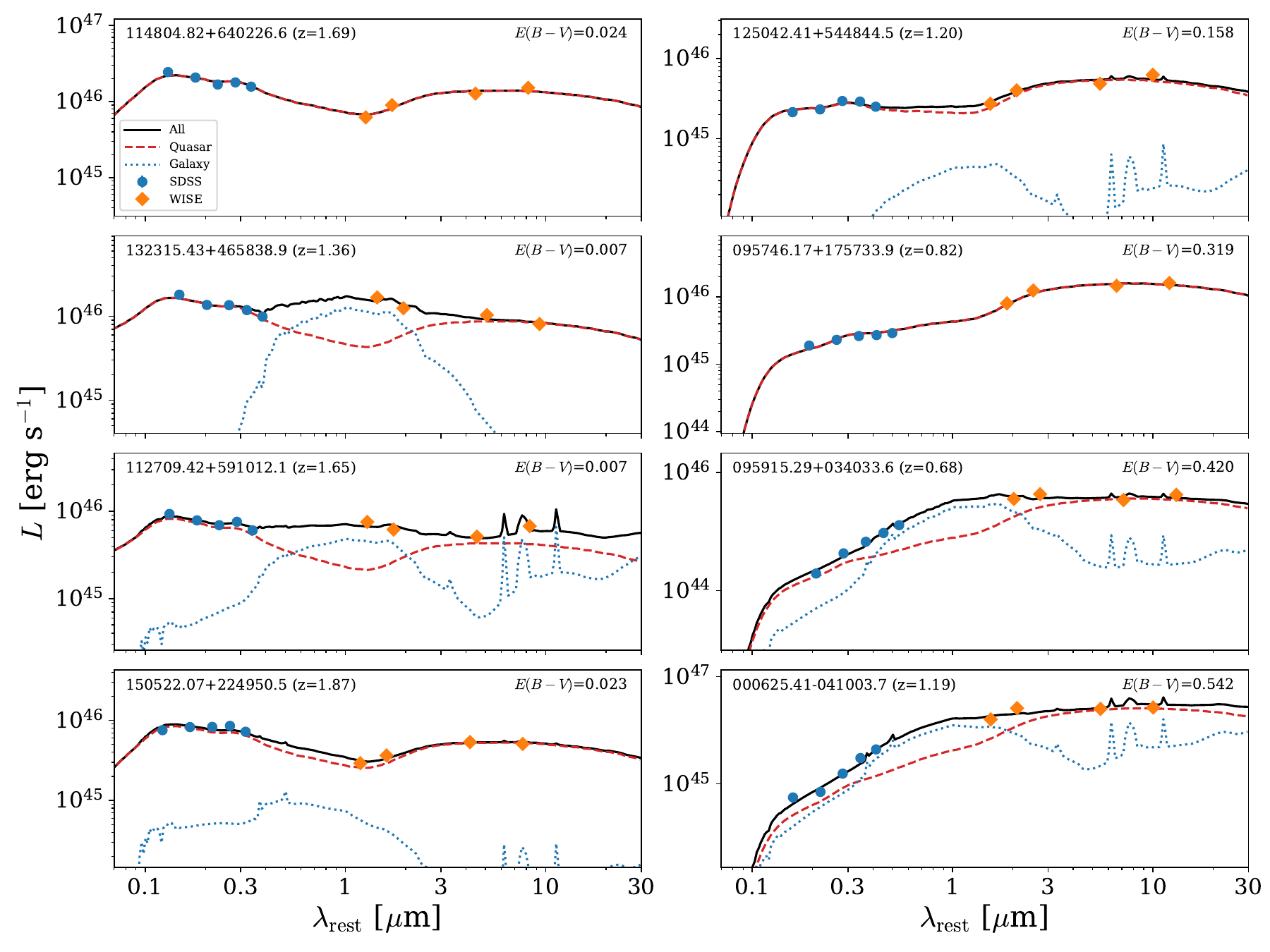}
\caption{
SED fitting results for selected objects from our sample in the rest-frame.
The blue and orange symbols represent the SDSS and WISE photometry, respectively.
The best-fit SED model is depicted by the solid black line, which includes contributions from the quasar component (red dashed line) and the galaxy component (blue dotted line). 
The left column panels show, from top to bottom, the cases largely contributed by quasar, elliptical, spiral, and irregular galaxy components.
The right column panels are for the various cases with significant reddening, where $\ebv>0.1$ mag.
\label{fig:sedfit}}
\end{figure*}

Among the DR14Q, we first select quasars detected in all of the SDSS ($u^*g'r'i'z'$) and AllWISE (W1W2W3W4) photometric data (9 bands in total) with a signal-to-noise ratio (S/N) criterion of S/N$>3$.
Note that the optical magnitudes are corrected for Galactic extinction \citep{Schlafly11,Paris18}.
While all these sources are brighter than the detection limits in each band in each survey, it is challenging to claim that they constitute a \emph{complete} quasar sample.
Considering the shape of the type-1 quasar SEDs \citep{Richards06b,Assef10,Krawczyk13}, it is worth noting that for the data we used, the optical imaging depths are 2-3 mag deeper than the MIR imaging depths.
In other words, the optical magnitudes corresponding to the MIR detection limits, as determined by the quasar SED templates, are 2-3 mag brighter than the optical detection limits.
Consequently, the selection process prior to this observation is more likely to favor the inclusion of quasars having significant MIR fluxes, particularly among those that are optically faint.
To avoid such a selection effect, we limit our sample to the complete quasar sample, following the criterion of $i'<19.1$ mag as noted by \cite{Richards06a}.
Under this criterion, the SDSS quasar survey is nearly complete ($\gtrsim 90$\,\%; see also \citealt{Vanden05}).

We employ the $L_{3000}$ and FWHM of \ion{Mg}{2}\,$\lambda 2798$ emission line (FWHM$_{\rm MgII}$) as proxies for BH activity.
Despite being secondary to primary indicators like optical luminosity ($L_{5100}$) and Blamer lines---and subject to an intrinsic dispersion of approximately 0.3 dex \citep{Ho12, Rakshit20}---these metrics facilitate the analysis of BH activity in quasars at higher redshifts through optical spectra available in the observed frame.
Given our objective to investigate the presence of red type-1 quasars up to cosmic noon ($z\sim2$), we opt to utilize the BH estimator based on $L_{3000}$ and FWHM$_{\rm MgII}$ as our principal estimator in this study.

We exclude the quasars for which the $L_{3000}$ values does not exist or the uncertainty in $L_{3000}$ exceeds  0.1 dex.
Note that the excluded sources by these $L_{3000}$ criteria occupy only a small fraction ($0.5\,\%$) of the total, showing $g'-i'$ color distributions consistent with those in the selected sample within the margin of error.
We also set a criterion of 2000\,km\,s$^{-1}< {\rm FWHM}_{\rm MgII}<$ 15000\,km\,s$^{-1}$ to exclude disk emitters and narrow line Seyfert 1 galaxies.
Note that \cite{Rakshit20} provides broad and narrow components separately, while we here use the broad component properties that are widely used for the $\mbh$ estimation (e.g., \citealt{Le20}).
Then, considering that the FWHM values obtained from the low-quality spectra are untrustworthy, we limit our sample to the quasars with a median S/N per pixel greater than 5 in the rest-frame range of 2700-2900\,$\rm\AA$, defined as {LINE\_MED\_SN\_MGII} in the \citeauthor{Rakshit20} catalog.
We also rejected the cases that the $\alplam$ values exceed the fitting range of $-5\leq\alplam\leq3$ in \cite{Rakshit20}, causing the $L_{3000}$ and \ion{Mg}{2} property measurements to be unreliable.

Then we restrict our sample to quasars at $0.68\leq z<2.20$.
The choice of the boundaries is primarily based on the edges of the redshift bins for QLF in \cite{Richards06a} to compare results parallelly.
Specifically, the edges of four redshift bins are [0.68, 1.06, 1.44, 1.82, 2.20], which are also used in the subsequent sections.
But we note that the completeness for the $L_{3000}$ and FWHM$_{\rm MgII}$ values are concerned as well.
In addition, the upper bound prevents a degenerative situation between the internal dust obscuration and the extraneous intergalactic medium attenuation by neutral hydrogen at higher redshift at shorter wavelength.
Specifically, the attenuation magnitude within the $u^*$-band experiences a dramatic increase for $z\gtrsim2.2$, as visually depicted in Figure 8 of \cite{Inoue14}.
All these criteria yield a preliminary selected sample size of 48,243.

\subsection{SED Fitting \label{sec:sedfit}}

The optical SEDs of quasars are dominated by quasar light from accretion disk, especially for high-luminosity ones, while host galaxy light occupies non-negligible contribution to the total SED of quasars at longer wavelengths \citep{Richards06b,Hickox18,Kim23}.
Hence, to measure the quasar luminosities in MIR precisely, it is required to precisely subtract the host galaxy light from their total luminosities.
In this section, we describe our methodology for decomposing the quasar and host galaxy components in the optical-to-MIR SEDs of our sample through SED fitting.
The SED fitting was performed in the similar way as \cite{Kim23}, so we briefly summarize the method below.

We utilize the quasar SED template of \cite{Richards06b}.
Our preference for this template stems from our utilization of the selection function established by \cite{Richards06a} in Section \ref{sec:impact}, which is determined from the quasar SED template.
This differs from \cite{Kim23} who used the more recent template of \cite{Krawczyk13}, derived from a larger sample.
However, as reported in \cite{Kim23}, the quasar SED template of \cite{Richards06b} shows no significant deviation from that of \cite{Krawczyk13} within the 0.1-20\,$\mu$m wavelength range, while that of \cite{Assef10} appears fainter at $1\,\mu$m  than the others.
Nonetheless, the measured monochromatic luminosities in MIR remain relatively unaffected by the choice of quasar SED template.
For galaxies, we use the three types of galaxy SED templates (elliptical, spiral, and irregular galaxies) of \cite{Assef10}.
The quasar SED template was interpolated to match the wavelength of the galaxy SED templates for unity.

The intrinsic quasar and galaxy SED templates, $f_0(\lambda)$, are reddened with a given $\ebv$ value as following:

\begin{equation}
f(\lambda) = 10^{- k(\lambda)E(B-V)/1.086} \times f_0(\lambda),\label{equ:ebv}
\end{equation}

\noindent where $k(\lambda)$ is the reddening law for starburst galaxies \citep{Calzetti00} that are free from the effect of the UV bump.
Note that the attenuation law in a functional form is extended beyond the original wavelength range (0.12\,$\mu$m $\leq \lambda_{\rm rest}\leq$ 2.2\,$\mu$m) for numerical calculation.
The reddened SED templates of quasar galaxies compose our SED model in the form of a linear combination:

\begin{equation}
f_{\rm comp}(\lambda)=C_{\rm Q}f_{\rm Q}(\lambda)+C_{\rm E}f_{\rm E}(\lambda)+C_{\rm S}f_{\rm S}(\lambda)+C_{\rm I}f_{\rm I}(\lambda),
\end{equation}

\noindent where $f_{\rm Q}$, $f_{\rm E}$, $f_{\rm S}$, and $f_{\rm I}$ are the reddened SED templates of quasar, elliptical, spiral, and irregular galaxies, respectively, as a function of wavelength, and $C_{\rm Q}$, $C_{\rm E}$, $C_{\rm S}$, and $C_{\rm I}$ are the coefficients associated with each respective template.
To conclude, there are five free parameters in our SED model: \{$\ebv$, $C_{\rm Q}$, $C_{\rm E}$, $C_{\rm S}$, $C_{\rm I}$\}.
In the fitting process, all the parameters are limited to have positive values, while we give a boundary condition for the extinction, $-1\leq \ebv\leq1$.
The negative $\ebv$ value is allowed for the quasars having bluer SED than the fiducial SED of \cite{Richards06b}, which has also been introduced in the previous works (e.g., \citealt{Glikman18}). 
The choice of the upper bound is based on careful consideration of the SED fitting results for our sample upon repetition, aiming to prevent a $\ebv$ value from reaching the boundary.

Using the above model, we performed the SED fitting for the optical ($u^*g'r'i'z'$) and MIR fluxes (W1W2W3W4) of our preliminary selected sample in Section \ref{sec:dr14q}, after they are deredshifted to the rest frame.  
In this process, we use the \texttt{MPFIT} package \citep{Markwardt09} for the Interactive Data Language (IDL) as in \cite{Kim23}.
For each band, 10\,\% of its flux is added in quadrature to the photometric uncertainty only in the SED fitting process to represent instrumental and calibration uncertainties, as in \cite{Boquien19}.
Figure \ref{fig:sedfit} shows the examples of the best-fit SED models with the photometric data.

\begin{figure}
\centering
\epsscale{1.2}
\plotone{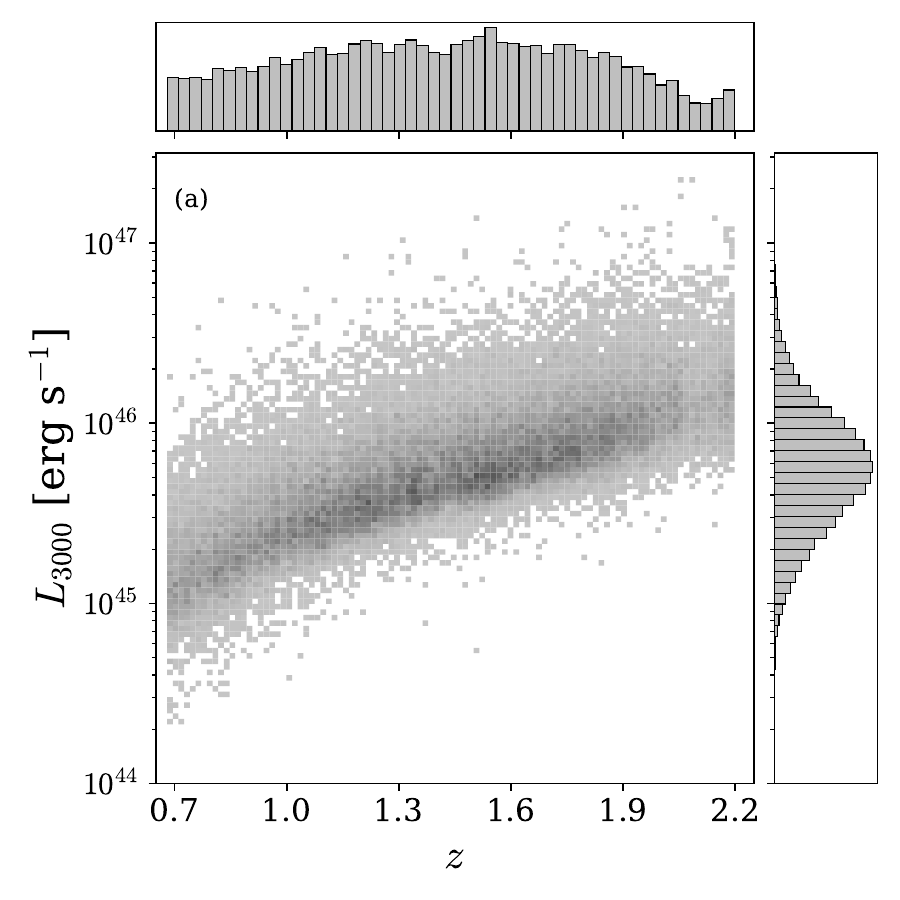}
\plotone{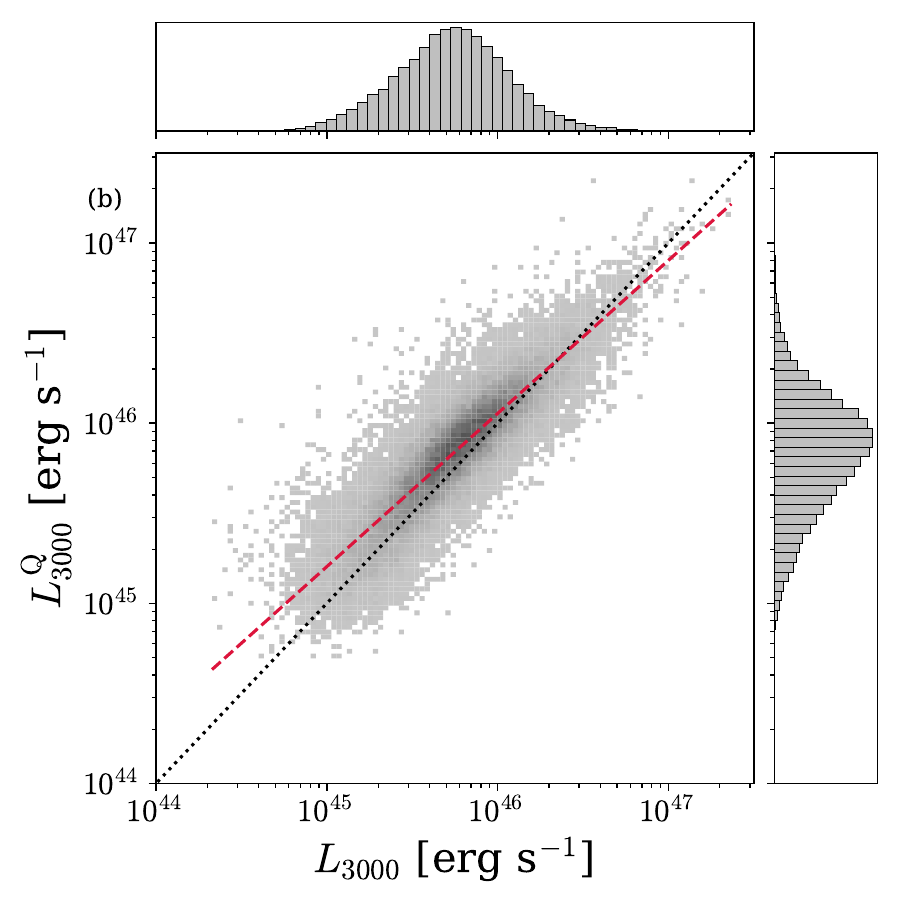}
\caption{
(a) The redshift and $L_{3000}$ distribution of our sample consisting of 48,241 quasars.
(b) Comparison of monochromatic luminosities from spectral fitting by \citeauthor{Rakshit20} (\citeyear{Rakshit20}; $L_{3000}$) and our SED fitting ($L^{\rm Q}_{3000}$).
The red dashed line shows the linear fit with a slope of 0.85, inferring the Baldwin effect.
The dotted line is an identical line.
\label{fig:sample}}
\end{figure}

While performing the SED fitting for the initial sample, a significant majority exhibited good fits, but there are several quasars with poor fitting results.
We define the fraction of light from quasar itself as $R_{\rm Q}(\lambda) = C_{\rm Q} f_{\rm Q} (\lambda)/f_{\rm comp}(\lambda)$.
Within this sample, there are two quasars with $R_{\rm Q} (3000\,{\rm\AA}) < 0.1$, one of which have $R_{\rm Q}(3000\,{\rm\AA})=0$.
While they could be considered galaxy-dominant, we exclude them as poor fitting results due to the presence of AGN features, such as broad components of H$\alpha$, H$\beta$, and/or \ion{Mg}{2} lines, in their spectra.
The remaining sample consists of 48,241 quasars, with the distribution of redshift and $L_{3000}$ shown in Figure \ref{fig:sample}(a).

From the SED fitting results, we estimate the monochromatic luminosity of \emph{quasar itself} by $L^{\rm Q}_{\lambda}=C_{\rm Q}f_{\rm Q}(\lambda)$.
The denoted superscript $(\rm Q)$ is to avoid a confusion with the monochromatic luminosity from spectral fitting by \cite{Rakshit20}.
For example, $L^{\rm Q}_{3000}$ is the monochromatic luminosity by a quasar alone at 3000\,$\rm\AA$ from the SED fitting.
As usual, the uncertainty of $L^{\rm Q}_{\lambda}$ is determined by the error propagation of $C_{\rm Q}$ and $f_{\rm Q}(\lambda)$; the former is determined from the SED fitting, while the latter is sourced from \cite{Richards06b}.
Likewise, the monochromatic luminosity of galaxy is estimated by $L^{\rm G}_{\lambda} = C_{\rm E}f_{\rm E}(\lambda)+C_{\rm S}f_{\rm S}(\lambda)+C_{\rm I}f_{\rm I}(\lambda)$, under the assumption that the galaxy SED is a combination of the galaxy components.

In Figure \ref{fig:sample}(b), we present a comparison between $L_{3000}$ and $L^{\rm Q}_{3000}$, revealing a tension, particularly noticeable in the faint regime.
This discrepancy can be attributed to the limitations of the quasar SED template we used, which integrates both emission and continuum line fluxes, implying that the estimation of $L^Q_{3000}$ is subject to contamination by the emission lines.
According to the Baldwin effect \citep{Baldwin81,Dietrich02}, the contamination by emission lines becomes increasingly significant as luminosity decreases.
In contrast, \cite{Rakshit20} derived $L_{3000}$ values after subtracting emission line components, reducing potential contamination by emission lines.
The observed power-law slope of 0.85 between the $L_{3000}$ and $L^{\rm Q}_{3000}$ values (red dashed line) supports this explanation.
Given these considerations, we proceed to adopt $L_{3000}$ as the definitive measure of UV luminosity for quasars in subsequent analyses, rather than $L_{3000}^{\rm Q}$.

\begin{figure*}
\centering
\epsscale{1.1}
\plotone{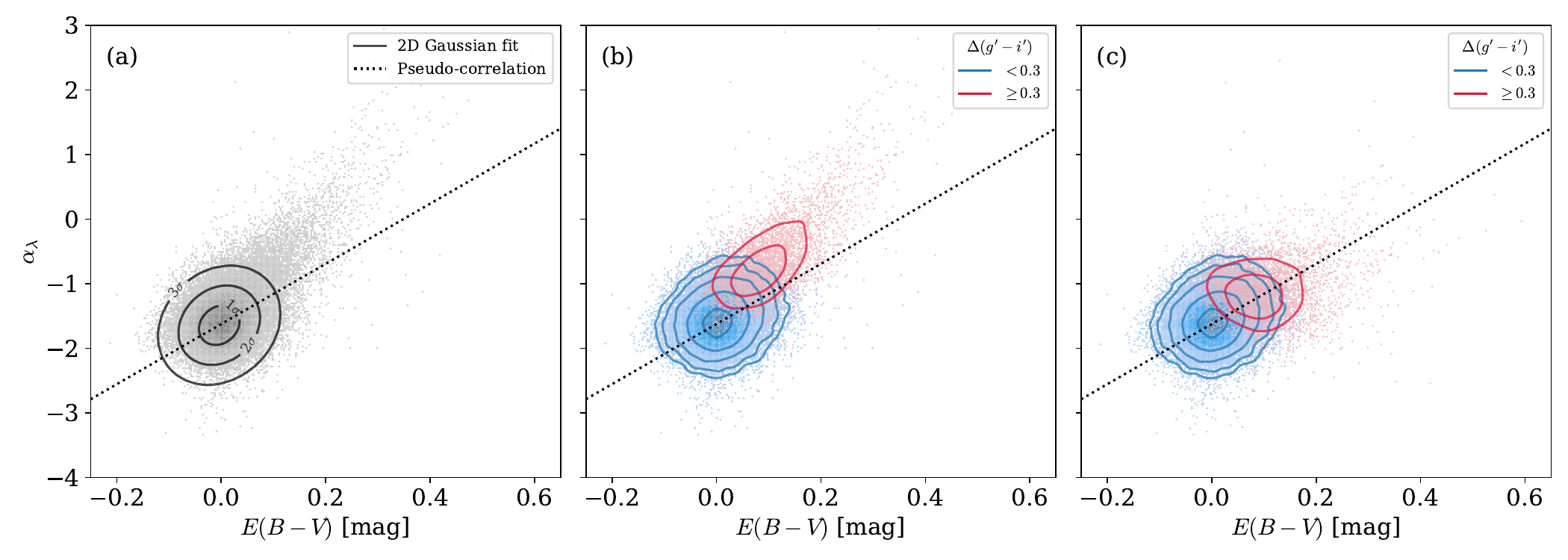}
\caption{
(a) Distribution of $E(B-V)$ and $\alpha_{\lambda}$. 
The contours show the 2D Gaussian fitting results.
The black dotted line illustrates the pseudo-correlation between the quantities, corresponding to the expected $\alpha_{\lambda}$ value as a function of $E(B-V)$ derived from the power-law function fitting to the quasar SED template of \cite{Richards06b}.
(b) Blue points represent data points with $\Delta(g'-i') < 0.3$, while red points represent data points with $\Delta(g'-i') \geq 0.3$.
Contours depict the Kernel Density Estimate (KDE) distributions, with levels of 5\,\%, 10\,\%, 20\,\%, 40\,\%, and 80\,\% of the maximum value, from outer to inner. 
(c) Similar to the panel (b), but displaying the distribution after the $E(B-V)$ correction for the red type-1 quasars.
\label{fig:ebvslope}}
\end{figure*}

\begin{figure}
\centering
\epsscale{1.1}
\plotone{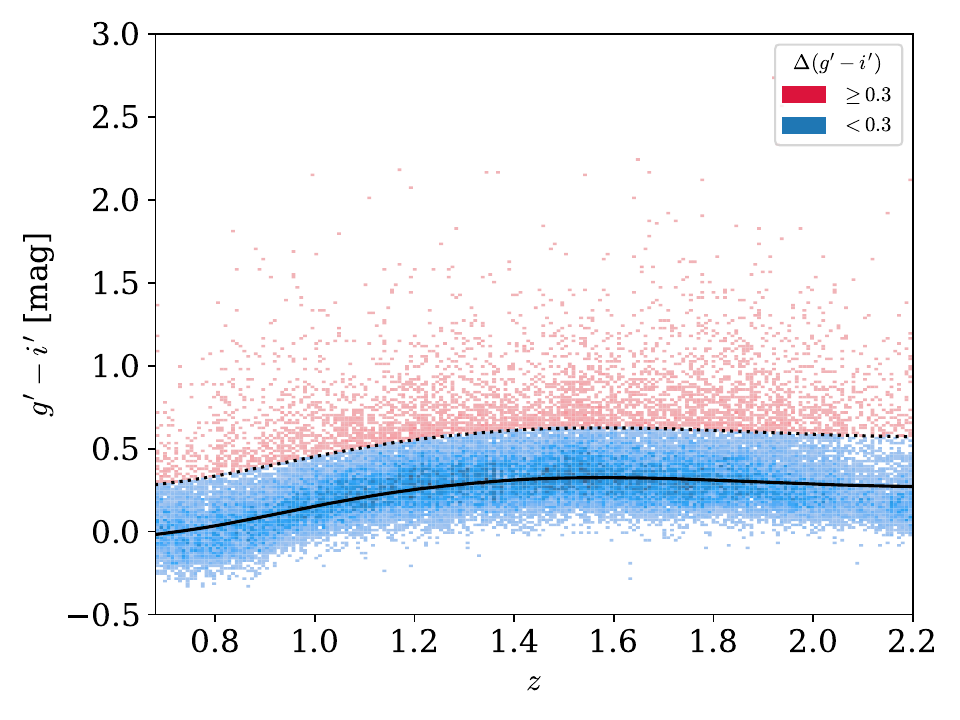}
\caption{
$g'-i'$ color distribution of our sample.
The black solid line represents the change in $g'-i'$ color of the quasar SED template we used \citep{Richards06b} as a function of redshift.
The dotted line indicates the $\Delta(g'-i')=0.3$ mag \citep{Lacy13}, a selection criterion for red type-1 quasars.
\label{fig:gicolor}}
\end{figure}

\subsection{Pseudo-correlation between $\ebv$ and $\alplam$ \label{sec:pseudo}}

The SED fitting results reveal that not all DR14Q is likely to be blue normal quasars, despite their classification as type-1 quasars characterized by strong and broad emission lines \citep{Kim23}.
In Figure \ref{fig:ebvslope}(a), we present the distribution of $\ebv$ and $\alplam$ in our sample, obtained through our SED fitting and the catalog by \cite{Rakshit20}, respectively.
A positive correlation, as anticipated by the reddening law, is confirmed by the 2D Gaussian fitting applied to the entire sample (black contours).
However, it is important to note that both the usage of fiducial quasar SED template with a fixed continuum slope and the inclusion of negative $\ebv$ values in our SED fitting process can lead to a pseudo-correlation between these parameters.
In other words, since the $\ebv$ values are determined from the tension between the fiducial quasar SED template and the observed SED shapes, the intrinsic variation in $\alplam$ (or quasar spectral shape) can create a pseudo-correlation between the two quantities.

To investigate the impact of $\ebv$ on $\alplam$, we conduct simulations using the quasar SED templates we employed \citep{Richards06b}.
Specifically, we apply reddening/dereddening with Equation (\ref{equ:ebv}) to the quasar SED templates for various $\ebv$ values.
In the spectral windows free from emission lines (1350-1365$\,\rm\AA$, 4200-4230$\,\rm\AA$; \citealt{Vanden01}), we fitted a power-law function given as Equation $(\ref{equ:fpl})$, to the adjusted SEDs to obtain the $\alplam$ for the given $\ebv$ values.
Caution is required when comparing the $\alplam$ values from \cite{Rakshit20} and this work, because the $\alplam$ measurement methods are slightly different.
The result is shown as the dotted line in Figure \ref{fig:ebvslope}.
While most data points align with the simulated $\alplam$, discrepancies emerge at $\ebv>0.1$ mag. 
This discrepancy could be attributed to red type-1 quasars, whose spectral shapes are affected by additional dust obscuration beyond the pseudo-correlation.

\section{Red Type-1 Quasars\label{sec:redtype1}}

\subsection{Classification by Color Excess}

The presence of the pseudo-correlation between $\ebv$ and $\alplam$ (Section \ref{sec:pseudo}) implies that a straightforward selection for dust-obscured quasars based on high $\ebv$ can lead to misclassification.
While the $\ebv$ values derived from the SED shape are used here, they can also be determined from the hydrogen line ratios (e.g., \citealt{KimIm18,Kim18}).
The $\ebv$ values obtained through these two methods exhibit a consistency with an rms dispersion of $\sim0.2$ \citep{Kim18}, indicating their limited accuracy.
Consequently, it becomes challenging to classify dust-obscured quasars solely based on the measured $\ebv$ values.

Instead, we introduce a color excess to classify red type-1 quasars in this study.
As inferred from the variation in $\alplam$, there is also an intrinsic variation in quasar colors.
While most of them are symmetrically distributed around the mean color of quasars, there are objects that exhibit a red color excess beyond this distribution, which are likely to be reddened by dust.
Both \cite{Richards03} and \cite{Lacy13} employed the $g'-i'$ color excess, denoted as $\Delta(g'-i')$, as a criterion for identifying red type-1 quasars.
$\Delta(g'-i')$ is defined as a difference between the observed $g'-i'$ color and the $g'-i'$ color of quasar SED template of \cite{Richards06a}; $ \Delta(g'-i') \equiv (g'-i')_{\rm obs} - (g'-i')_{\rm template}$.
Figure \ref{fig:gicolor} shows the $g'-i'$ color distribution of our sample.
Taking account of the change in $g'-i'$ color of the quasar SED template of \cite{Richards06b} according to redshift (black solid line), we obtain the $\Delta(g'-i')$ values of our sample.
\cite{Lacy13} defined red type-1 quasars as those with $\Delta(g'-i')\geq 0.3$ mag (dotted line).
We apply this criterion to our sample, resulting in the division of type-1 quasars into two subsamples: blue ($\Delta(g'-i')<0.3$) and red ($\Delta(g'-i')\geq0.3$), shown in corresponding colors in Figures \ref{fig:ebvslope}(b), (c) and \ref{fig:gicolor}. 
The numbers of them are 43,631 and 4610, respectively.

In Figure \ref{fig:ebvslope}(b), the blue type-1 quasars exhibits characteristics resembling a 2D Gaussian distribution, skewed toward the pseudo-correlation mentioned earlier. 
It has a median and 1$\sigma$ standard deviation of slope $\alplam=-1.58^{+0.32}_{-0.29}$, consistent with both the slope of fiducial quasar SED template ($\alplam=-1.62$) and the canonical values from quasar UV spectra \citep{Vanden01, Telfer02, Lusso15, Selsing16}.
On the other hand, the red type-1 quasars becomes more prominent as $\ebv$ values increase and reveals noticeable discrepancies between data points and the pseudo-correlation.
They exhibits redder slopes than the pseudo-correlation, with $\alplam=-0.72^{+0.55}_{-0.37}$, which may be the result of dust obscuration.

Under the assumption that the intrinsic slope of red type-1 quasars is consistent with the slope of the fiducial quasar SED template, the dereddened slopes are in line with the pseudo-correlation, shown in Figure \ref{fig:ebvslope}(c) as opposed to the original values in the panel (b).
Note that this process is similar to how we derived the pseudo-correlation mentioned earlier.
The alignment with the pseudo-correlation, instead of being centered on the fiducial slope value ($\alplam\sim-1.6$) of the quasar SED template, implies that the redder colors and slope are indeed the result of dust obscuration, irrespective of the intrinsic variations in quasar spectral shape.

In order to confirm that the red $g'-i'$ color of the red subsample originates from the dust extinction, additional tests using line ratio and X-ray hardness are needed.
However, such analyses are beyond the scope of this study.
In the following sections, we proceed with the assumption that the red color of these red type-1 quasars is a result of dust extinction (e.g., \citealt{KimIm18,Kim23}), and we correct their intrinsic luminosities using the measured $\ebv$ values.

\begin{figure*}
\centering
\epsscale{1.2}
\plotone{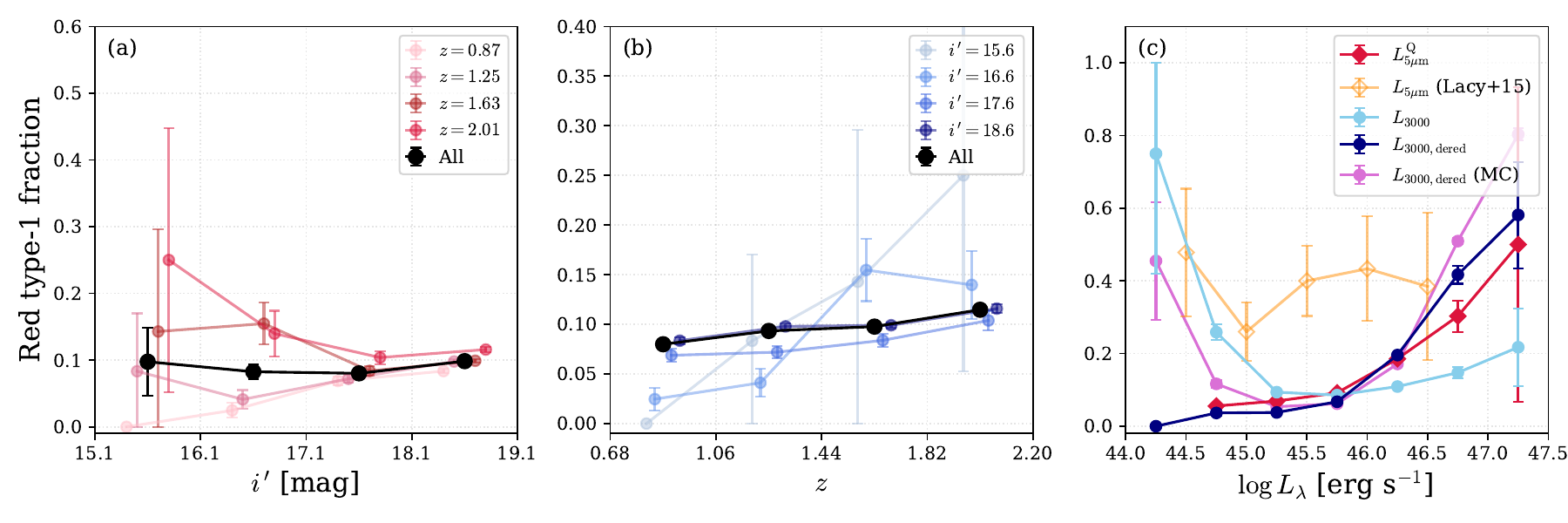}
\caption{
(a) Fraction of red type-1 quasar as a function of $i'$-band magnitude.
The colored symbols represent the fraction at different redshift bins, while the black symbols are for all sample.
For visibility, the colored symbols are mildly shifted in x-axis.
The vertical dotted lines indicate the edges of the magnitude bins.
(b) Redshift evolution in the red type-1 quasar fraction (black symbols).
The colored symbols represent the fraction at different magnitude bins.
(c) Fraction of red type-1 quasar with respect to monochromatic luminosities.
The red diamonds represents the fraction concerning $L_{5 \mu m}$, while the open orange diamonds are that from an MIR-selected quasar survey \citep{Lacy15}.
The skyblue circles are the fraction as functions of $L_{3000}$, and the navy circles are the case for the dereddened $L_{3000}$ ($L_{3000, \rm{dered}}$)with $\ebv$ values.
The orchid circles are the case for $L_{3000, \rm{dered}}$ from the Monte Carlo simulation.
\label{fig:magzlum}}
\end{figure*}

\subsection{Fraction of Red Type-1 Quasars}

It is worth noting that the fraction of red type-1 quasar is 9.6\,\% (4610/48,231), indicating a non-negligible fraction of obscured quasars within the DR14Q dataset.
This result is consistent with that of \cite{Kim23}, albeit with minor differences in the used quasar SED template \citep{Krawczyk13} and the criteria for selecting red type-1 quasars.

Figure \ref{fig:magzlum} shows the changes in the fraction of red type-1 quasars with respect to $i'$-band magnitude (a), redshfit (b), and monochromatic luminosities (c).
We found that the fraction tends to be higher at higher redshift and brighter $i'$-band magnitude, although the uncertainties are relatively large due to the small number statistics.
This result differs from that of the red type-1 quasars selected through fixed color criteria (e.g., $R-K>4$ and $J-K>1.8$; \citealt{Glikman12}) regardless of redshift (but note that it is well understood that these previous results are affected by selection effects; \citealt{Glikman04,Glikman12}).

In panel (c), we present the red type-1 quasar fraction relative to their monochromatic luminosities.
It shows an upsurge correlated with increasing MIR luminosity at 5$\,\mu$m ($L^{\rm Q}_{5\mu m}$) from the SED fitting.
This trend contrasts with the typically observed decline in obscured quasar fraction with higher luminosity reported in the literature \citep{Ueda14,Aird15,Lacy15,Glikman18,Vijarnwannaluk22}.
However, it is crucial to note that our analysis focuses solely on type-1 quasars, presenting a different trend as in the results from the MIR-selected surveys \citep{Lacy15,Glikman18}.
The rising fraction aligns with expectations from our $i'$-band magnitude-limited dataset and the $\Delta(g'-i')$ color criterion for selecting red type-1 quasars, in terms of completeness.
Our fraction significantly undercuts the 40\,\% red type-1 quasar fraction seen in MIR-selected surveys (open orange diamonds; \citealt{Lacy15}) but parallels their results at  $L^{\rm Q}_{5 \mu m}=10^{47}$ erg s$^{-1}$, which has high completeness in MIR.
This convergence strengthens the validity of our observations and supports the stated assertion.
Keep in mind that both $L^{\rm Q}_{5 \mu m}$ and $L_{5 \mu m}$ of \cite{Lacy15} are derived from the SED fitting applied to broadband photometry, which smears out the flux contamination by emission lines when comparing them.
We also note that the change in trend due to redshift is insignificant.
In conclusion, there remains potential for red type-1 quasar fraction to increase if optically fainter samples are included\footnote{Expanding the sample beyond $i'<19.1$ mag suggests a potential increase in the fraction to approximately 13\,\%. However, the completeness issues in optical of this broader dataset preclude a detailed discussion within this context.}.

Note that we only present the case of $L_{5 \mu m}$ for comparison with \cite{Lacy15}.
Considering the negligible impact of dust obscuration in NIR, utilizing the fiducial quasar SED template allows us to draw parallels with red type-1 quasar fractions at different NIR wavelengths (e.g., $L_{3.4 \mu m}$ \& $L_{4.6 \mu m}$ as used in \citealt{Kim23}), which demonstrate similar trends to those observed in $L_{5 \mu m}$.

We also find changes in the fraction as a function of $L_{3000}$ (blue circles), which naturally exhibits a higher fraction at the faint ends.
However, upon dereddening the $L_{3000}$ values with their corresponding $\ebv$ values ($L_{3000, \rm{dered}}$), the fraction (navy circles) increases notably at the bright ends, mirroring the trend seen with $L_{5 \mu m}$.
Specifically, the number of quasars with $L_{3000}>10^{46}$ erg s$^{-1}$ rises from 9926 to 11,244 (a 13\,\% increase), while the count of red type-1 quasars sees a more significant surge, from 1111 to 2429 (a 219\,\% increase).
This elevated fraction is expected under the assumption that intrinsic quasar SEDs align with the fiducial quasar SED templates we employed.
However, as discussed in Section \ref{sec:sedfit}, quasars exhibit diverse SED shapes, introducing a systematic uncertainty in $\ebv$ estimation in our work.

To address this uncertainty, we consider a systematic uncertainty in $\ebv$ of 0.05 mag, based on the 2D distribution of the $\ebv$-$\alplam$ plane (see Figure \ref{fig:ebvslope}).
We conduct a Monte Carlo simulation, generating 10,000 sets of the red type-1 fraction as a function of $L_{3000, \rm{dered}}$.
The resulting fraction, shown as the gray circles in Figure \ref{fig:magzlum}, closely aligns with the result obtained without considering the systematic uncertainty in $\ebv$.
Note that the higher fractions of the brightest bins are attributed to the small sample sizes, which vary significantly from one bin to another, differing by an order of magnitude.
This change in red type-1 quasar fraction suggests the presence of numerous quasars whose observed UV luminosities in the rest frame ($L_{\rm UV}$) are underestimated, potentially impacting the statistics related to bright quasars based on $L_{\rm UV}$.
We note that according to the fiducial quasar SED template \citep{Richards06a},  $L_{3000}$ is an order of magnitude brighter than $L_{5 \mu m}$.

\section{Impact on $L_{\rm UV}$-related Quasar Statistics\label{sec:impact}}

Our finding of non-negligible fraction of red type-1 quasar suggests that the dust obscuration is required to be concerned when studying fundamental properties of SDSS quasars based on $L_{\rm UV}$.
In this section, we examine the impact of the prevalence of red type-1 quasars on the quasar demographics and SMBH activities.

\subsection{Quasar Luminosity Function\label{sec:qlf}}

The change in $\alplam$ due to dust obscuration in Section \ref{sec:redtype1} implies that the observed luminosities of the red type-1 quasars are also underestimated in the same manner, especially at shorter wavelengths.
In practice, the number of intrinsic luminous quasars increases after dereddening.
Therefore, we  estimate QLF to assess whether its shape changes significantly when considering the presence of red type-1 quasars.

QLF is typically derived from a volume-limited sample with meticulous consideration of various incompleteness in survey process.
In particular, selection function, which includes photometric, spectroscopic, and selection completeness, is widely used to correct the number of the observed ones (e.g., \citealt{Richards06a,Ross13,Palanque13,Palanque16,Kim15,Kim20,Kim22}).
While our dataset has 48,231 quasars including 9.6\,\% being red Type-1 quasars, they are not uniformly selected, making them inappropriate for calculating QLF while considering such selection effects.
Instead, we opt to use the homogeneous statistical sample of \cite{Richards06a} by matching it to our sample.

\begin{figure}
\centering
\epsscale{1.2}
\plotone{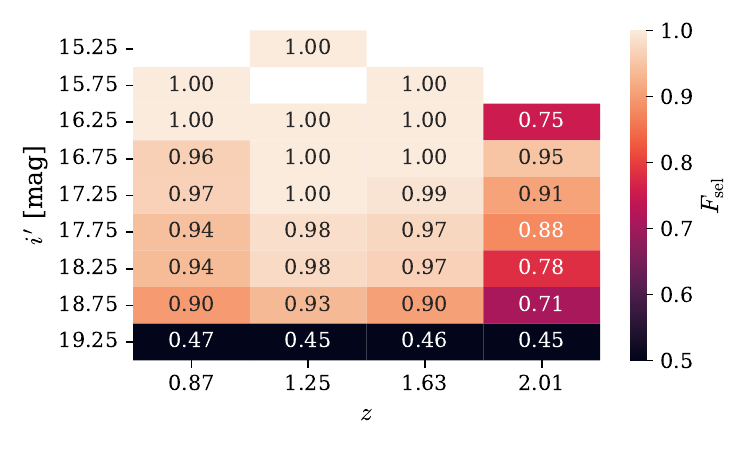}
\caption{
Selection completeness $F_{\rm sel}$ of our sample compared to the homogeneous quasar sample of \cite{Richards06a}.
\label{fig:comp}}
\end{figure}

\cite{Richards06a} utilized a sample of 15,343 SDSS DR3 quasars over 1622 deg$^{2}$.
They also provide absolute magnitudes in the rest-frame $i'$-band at $z=2$ ($M_{i'}(z=2)$) of the sample and values of the selection function ($F_{\rm R06}$).
Following \cite{Kulkarni19}, we convert $F_{\rm R06}$ given in observed $i'$-band magnitude to absolute magnitude ($M_{i'}(z=2)$), including $K$-correction provided by \cite{Richards06a}.
For the construction of binned QLFs, \cite{Richards06a} employed manually optimized redshift bins.
As in Section \ref{sec:redtype1}, we use four redshift bins that align with our main sample.
There are 10,175 SDSS DR3 quasars with $i'<19.1$ mag at $0.68\leq z<2.20$, 8458 of which are matched to our main sample.
The omitted sample is due to our selection criteria in Section \ref{sec:sample}, which can give additional selection effects.
Figure \ref{fig:comp} shows the selection completeness ($F_{\rm sel}$) of our sample as functions of $i'$-band magnitude and redshift, compared to the \cite{Richards06a} sample.
We include this selection effect in the QLF calculation according to the sources' $i'$-band magnitudes.

There are 695 red Type-1 quasars in the matched sample, showing a fraction of $8.2\,\%$ (695/8458) comparable to that of our main sample.
Similar to the work in Section \ref{sec:redtype1}, we deredden $M_{i'}(z=2)$ of the red Type-1 quasars according to their $\ebv$ values.

Then we calculate binned QLFs as in \cite{Page00}. 
The specific comoving volume of a bin ($V_{\rm bin}$) is defined as

\begin{equation}
V_{\rm bin} =\iint F_{\rm R06}(M,z) \frac{dV_{c}}{dz}dz\,dM,
\end{equation}

\noindent
where $dV_{c}/dz$ is the comoving element of the survey area (1622 deg$^{2}$), and $M$ is the absolute magnitude.
The same magnitude bins with a bin size of 0.3 mag as in \cite{Richards06a} are also used.
When evaluating the integral, we simply sum the $F_{\rm R06}$ elements within the magnitude and redshift range without any interpolation as in \cite{Kim20,Kim22}.
Then, the binned QLF is given as

\begin{equation}
\Phi_{\rm bin} (M,z) = \frac{N_{\rm cor}}{V_{\rm bin}},
\end{equation}

\noindent where $N_{\rm cor}$ is the number of quasar in the given bin, corrected by $F_{\rm sel}$; $N_{\rm cor} = \sum^{N}_{i} (1/F_{{\rm sel},i})$, where $N$ is the real number of quasars in the bin and $F_{{\rm sel},i}$ is the selection completeness for $i$-th source.

\begin{figure*}
\centering
\epsscale{1.15}
\plotone{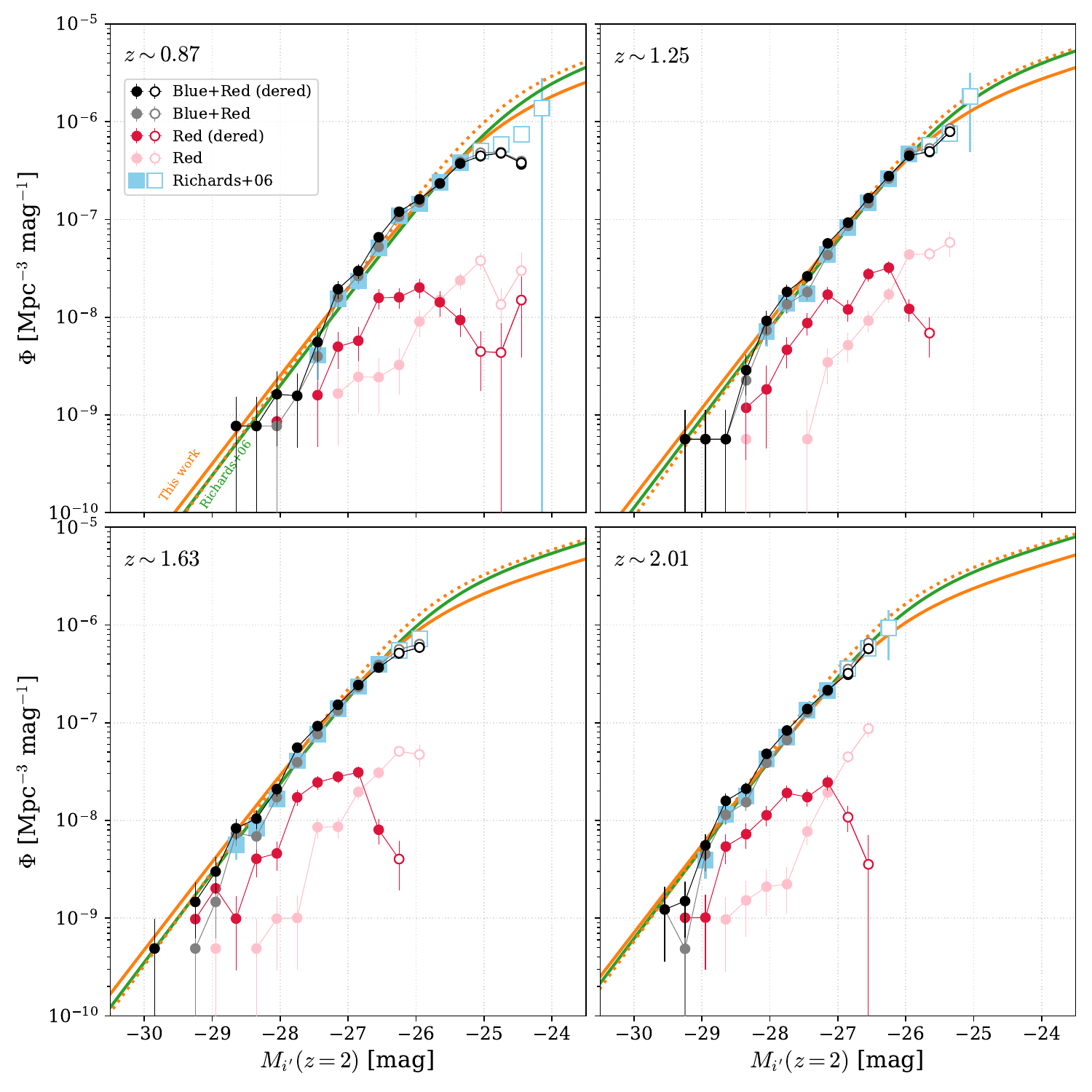}
\caption{
Differential QLFs at different redshift bins.
The black and gray circles are the QLF of our sample with and without the $\ebv$ correction for the red type-1 quasars, respectively.
Similarly, the red and pink circles are the QLF only for the red type-1 quasars.
The sky-blue squares represent the QLF of \cite{Richards06a}.
The filled and open symbols indicate complete and incomplete bins, respectively, as defined by \cite{Richards06a}.
The orange and green solid lines are the best-fit parametric QLFs for our dereddened sample and \cite{Richards06a} original sample within the complete bins, respectively.
The orange dotted lines represent our parametric QLFs with a strict luminosity cutoff (see text).
\label{fig:qlf}}
\end{figure*}

In Figure \ref{fig:qlf}, the gray circles represent the QLFs derived from our sample, which are in line with those presented by \citeauthor{Richards06a} (\citeyear{Richards06a}; blue squares).
Note that the filled and open symbols are for the complete and incomplete bins, respectively, defined by \cite{Richards06a}.
Additionally, the subset of red type-1 quasars is shown as pink circles.
When we deredden the magnitudes of the red type-1 quasars according to their $\ebv$ values, their QLFs shift toward the bright side (red circles).
This naturally results in an increase in the overall distribution at the bright end (black circles).
However, these differences are minor and become barely noticeable in higher redshift bins.

To see quantitatively, we fit the binned QLFs using a broken power-law function that is canonically used for estimating parametric QLFs at various redshifts (e.g., \citealt{Kulkarni19,Shen20,KimIm21}).
The parametric QLF is given as,

\begin{equation}
\Phi_{\rm par} = \frac{\Phi^{*}}{10^{0.4(\alpha+1)(M-M^{*})}+10^{0.4(\beta+1)(M-M^{*})}},
\end{equation}

\noindent where $\Phi^{*}$ is the normalization factor, $M^{*}$ is the break magnitude, and $\alpha$ and $\beta$ are the faint-end and bright-end slopes, respectively.
Our quasars are biased toward bright ones so we fixed the faint-end slope to $\alpha=-1.9$, sourced from the UV QLF derived by \cite{Kulkarni19}.
After cosmic noon, the QLF evolution can be approximated by a pure luminosity evolution (PLE; \citealt{Boyle00,Croom04,Croom09,Ross13}).
Therefore, we include redshift evolution on the break magnitudes as in \cite{Richards06a}:

\begin{equation}
M^{*} = M^{*}_{0} + m_1 \xi + m_2 \xi^2 + m_3 \xi^3,
\end{equation}

\noindent where $M^{*}_0$ is set to $-26$ mag, $\xi$ is the redshift scaling factor based on the reference redshift of $z_{\rm ref}=2.45$, which is defined as

\begin{equation}
\xi = \log \left( \frac{1+z}{1+z_{\rm ref}} \right),
\end{equation}

\noindent and $m_1$, $m_2$, $m_3$ are the polynomial coefficients of $\xi$.
Consequently, there are five free parameters in the $\Phi_{\rm par}$ evaluation ($\Phi^{*}$, $\beta$, $m_1$, $m_2$, $m_3$), and we fit $\Phi_{\rm bin} (M,z)$ with this broken power-law function using Python \texttt{curve\_fit} package.
It is important to note that we only fit the complete bins (filled symbols), defined by \cite{Richards06a}.

The orange and green lines in Figure \ref{fig:qlf} represent the best-fit results of our QLF and that of \cite{Richards06a}, respectively.
As anticipated from the binned QLF, the dereddening effect increases the number density of quasars at the bright ends, but there is no significant difference between the bright-end slopes: $\beta=-3.26\pm0.06$ and $-3.30\pm0.04$, respectively.
Given the optical magnitude cutoff ($i'<19.1$ mag) in the selection stage, however, low-luminosity red quasars could be missed (see Section \ref{sec:seleffect}), which could lead to an artificial flattening of the dereddened QLFs at the faint regime.
Indeed, the turnoff of number density exhibits even within the complete bins for the dereddened QLF of red type-1 quasars.
If we fit the QLF without two faintest complete bins in each redshift bin, where the turnoff becomes significant, we obtain the parametric QLF (orange dotted line) and its bright-end slope of $\beta=-3.41\pm0.06$.
The shape of this QLF is in line with that of \cite{Richards06a} at the bright regime, inferring the minor impact of the presence of red type-1 quasars on the QLF as well.

Since the $\beta$ estimation is correlated with the other parameters, we fix the luminosity evolution parameters ($m_1$, $m_2$, $m_3$) to those of the best-fit results of \cite{Richards06a}.
The change gives a slightly flatter bright-end slope of $\beta=-3.20\pm0.04$.
However, this difference of $\sim0.1$ is within the margin of errors in local estimations within the redshift bins (e.g., \citealt{Kulkarni19,Shen20}).
We also test the inclusion of the redshift evolution of $\beta$ in a similar manner as in \citeauthor{Shen20} (\citeyear{Shen20}; $\beta = \beta_{0}+b_1 \xi$), but there are not much differences from the above results.
Therefore, we conclude that the prevalence of red type-1 quasars only causes minor changes in quasar demography statistically.

Our study reveals that the number density of red type-1 quasars remains below that of typical blue (or unobscured) type-1 quasars even after applying the $\ebv$ correction. This finding deviates slightly from \cite{Banerji15} on extremely obscured quasars with $0.5<\ebv<1.5$. This is likely because our sample, derived predominantly from SDSS type-1 quasars chosen for their optical characteristics, underrepresents the most heavily obscured quasars which could impact the overall density distribution. Within our sample, only nine possess $\ebv$ values over 0.5, with the highest recorded at $\ebv=0.60$. Anticipating more obscured quasars from future mission like Spectro-Photometer for the History of the Universe, Epoch of Reionization, and Ices Explorer (SPHEREx; \citealt{Dore14}), it is crucial to consider their potential impact on the QLF.
Nevertheless, our findings reinforce the conclusion that a notable fraction of quasars are obscured \citep{Assef15,Banerji15,Kim23}, even within optically selected type-1 quasar samples. 
Additionally, their proportion increases with luminosity as well (see Figure \ref{fig:magzlum}). This underscores the importance of carefully accounting for such obscuration when empirically simulating QLF evolution (e.g., \citealt{Conroy13,Ren20,KimIm21}) to accurately reproduce observed QLF trends.

\subsection{$\mbh$ and $R_{\rm Edd}$ Estimations \label{sec:mbh}}

\begin{figure}
\centering
\epsscale{1.2}
\plotone{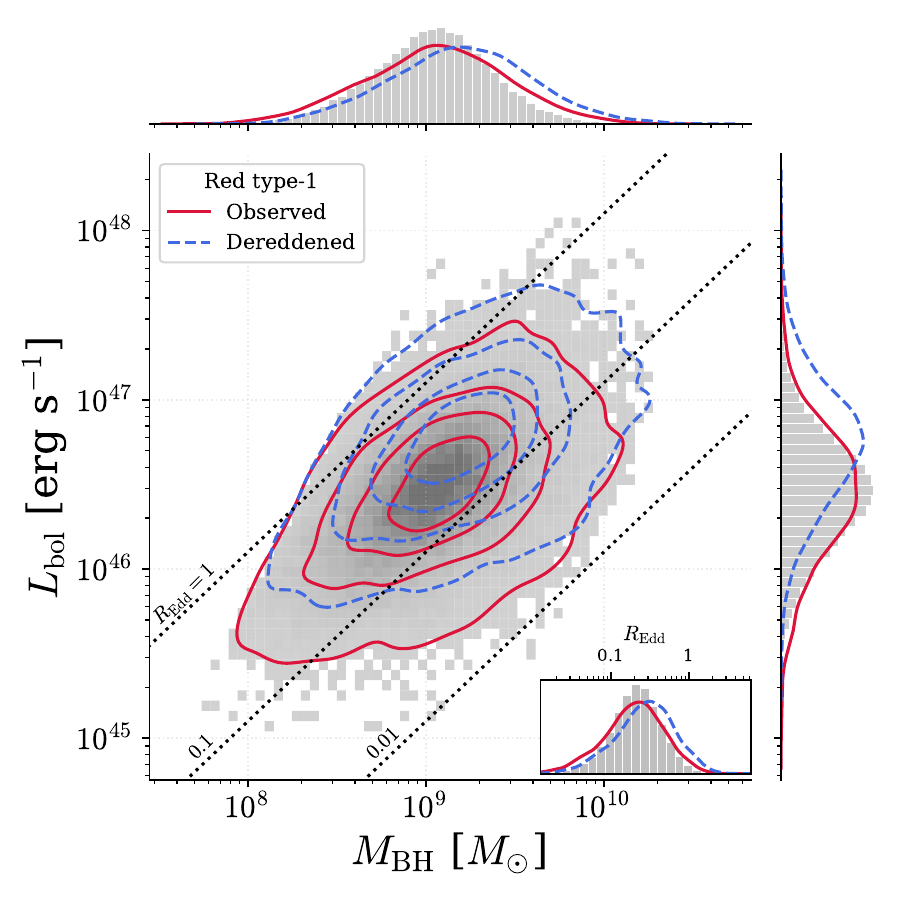}
\caption{
$\mbh$-$\lbol$ distributions for our main sample (gray histograms).
The red contours represent the distribution of the red type-1 quasars, whereas the blue dashed contours show the same distribution after applying the $\ebv$ correction.
The black dotted lines represent the constant $\redd$.
The inset shows $\redd$ distributions.
\label{fig:mbhdist}}
\end{figure}

There have been various attempts to measure the mass of the SMBHs.
For example, reverberation mapping is known to be a direct method to estimate the black hole mass ($\mbh$) by measuring the AGN structure using time delay between different spectral components (e.g., \citealt{Shen19,Shen23}).
But this requires long-term monitoring and therefore is not suitable for large samples.
Instead, previous studies have been measuring $\mbh$ from the spectral properties such as luminosity and FWHM of emission lines, under the assumption of rotating motions in broad-line region (BLR) gas around the BH with the size-luminosity relations (e.g., \citealt{Bentz09,Bentz13}).
This is known as the single epoch method, and \cite{VO09} presents a functional form for $L_{3000}$ and \ion{Mg}{2} line as follows:

\begin{equation}
\begin{aligned}
\log\left(\frac{\mbh}{M_{\odot}}\right)=&6.86+0.5\log\left(\frac{L_{3000}}{10^{44}~{\rm erg~s}^{-1}}\right)\\
&+2\log\left(\frac{{\rm FWHM_{\rm MgII}}}{1000~{\rm km~s}^{-1}}\right).\label{equ:mbh}
\end{aligned}
\end{equation}

\begin{figure}
\centering
\epsscale{1.2}
\plotone{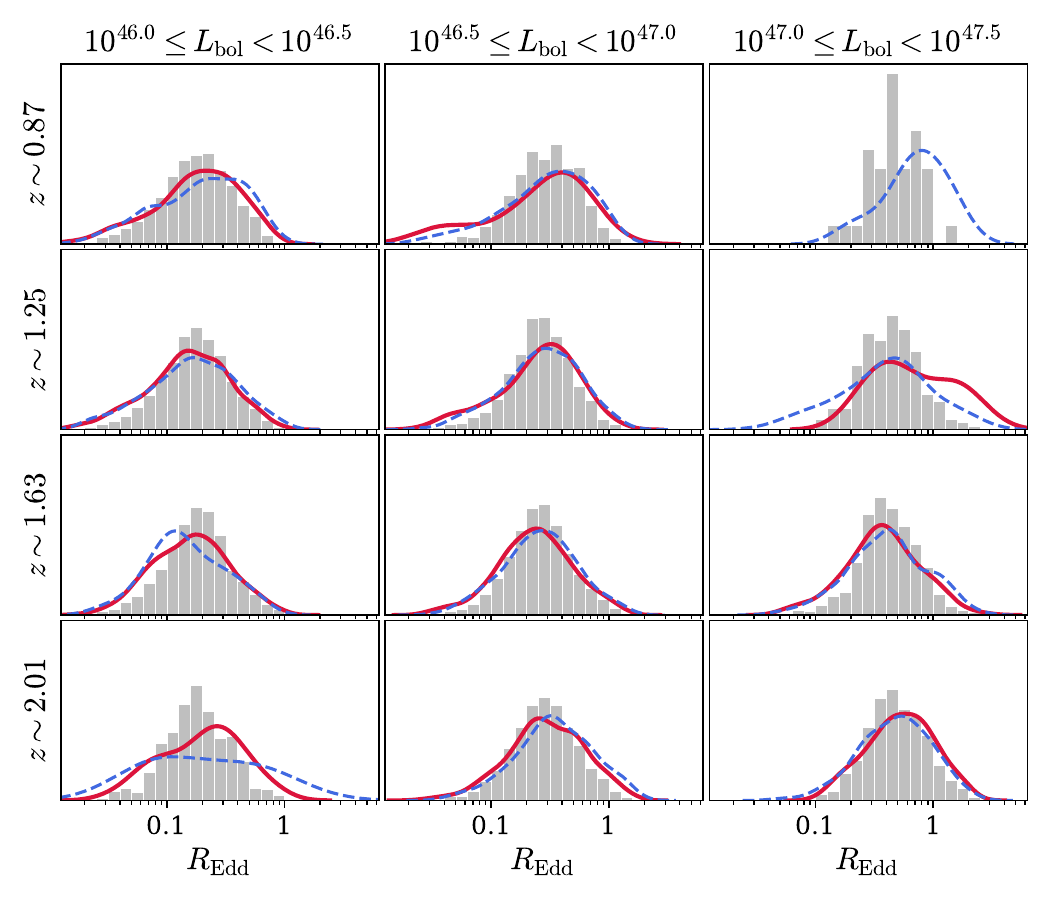}
\caption{
$\redd$ distributions in various redshift and $\lbol$ bins.
The gray histograms are for our main sample, while the red solid and blue dashed lines represent the distributions of red type-1 quasars before and after the $\ebv$ correction, respectively.
Note that in the top right bin, there is only one red type-1 quasar without the $\ebv$ correction, giving no distribution.
\label{fig:reddmatrix}}
\end{figure}

Using this equation, we estimate the $\mbh$ values for our main sample.
In Figure \ref{fig:mbhdist}, we show the $\mbh$ distribution of our sample alongside the bolometric luminosity derived by $\lbol=5.15\times L_{3000}$ \citep{Shen11}.
Note that the bolometric correction factor is determined from the quasar SED template of \cite{Richards06b}.
The MIR-based $\lbol$ estimator introduced by \cite{Kim23} is applicable up to $z\sim2.5$ \citep{Kim24b}.
However, for our analysis, we employ the $L_{3000}$-based estimator to examine the change in $\mbh$ following the $\ebv$ correction.
On the $\mbh$-$\lbol$ plane, the subset of red type-1 quasars (red contours) has a distribution consistent with the whole sample (gray histograms).
If we use $L_{3000, \rm{dered}}$ when estimating both $\lbol$ and $\mbh$, however, the distribution shift toward higher $\lbol$ and $\mbh$ values, shown as the blue dashed contours.
This is a natural behavior, considering Equation (\ref{equ:mbh}), i.e., change in $L_{3000}$ can affect the $\lbol$ and $\mbh$ estimation by factors of 1 and 0.5 in a logarithmic manner, respectively.
Indeed, the shifts between two distributions in $\lbol$ and $\mbh$ directions are 0.28 and 0.14 dex, respectively.
Similarly, the $\redd$ distribution, which is proportional to $\lbol/\mbh$ (or $\propto (L_{3000})^{1/2}$), also shifts accordingly.
The inset panel of Figure \ref{fig:mbhdist} shows the $\redd$ distributions, revealing a shift in the peak value of 0.14 dex.

Given the bias in our sample toward brighter and more massive quasars, it is challenging to conclude that every red type-1 quasar is in the blowout phase preceding normal type-1 quasars.
In Figure \ref{fig:reddmatrix}, we compare the distributions of $\redd$ for both the $z$ and $\lbol$ matched samples, showing no significant differences before and after the $\ebv$ correction.
Rather, their distributions are similar to those of main sample including normal type-1 quasars, even after the $\ebv$ correction.
This suggests the possibility that red type-1 quasars may possess intrinsically red slopes without substantial dust obscuration.
While hydrogen line ratios can typically be used to gauge dust extinction beyond the broad-line region of quasars \citep{Rose13,KimIm18}, our sample lacks estimations of hydrogen lines due to their considerable redshifts. 
Future infrared surveys such as SPHEREx \citep{Dore14} hold promise for estimating the hydrogen line ratios of our sample, which could provide insights into their nature and quasar evolution.

Another promising point is that the $\mbh$ estimates of individual red type-1 quasars have been underestimated by previous studies.
Unfortunately, the systematical uncertainty in the $\mbh$ estimation is known to be about 0.4 dex in the sense of the single epoch method \citep{Vestergaard06,Shen08,Shen13}.
Only 1.9\,\% of the red type-1 quasars show change in $\mbh$ greater than 0.4 dex, meaning that we cannot confidently say that the remaining sources' $\mbh$ values are truly underestimated.

\section{Discussion \label{sec:discussion}}

\begin{figure}
\centering
\epsscale{1.2}
\plotone{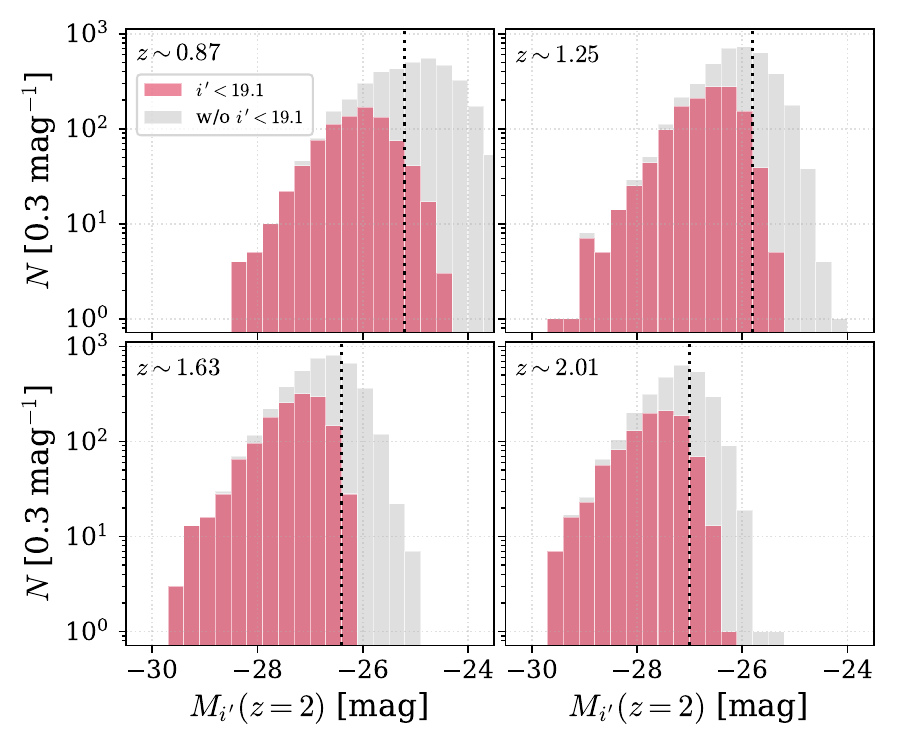}
\caption{
Differential number counts of red type-1 quasars after the $\ebv$ correction.
The red histograms are for our main sample with optical magnitude cutoff with $i'<19.1$ mag, whereas the gray histograms represent the extended sample without this cutoff.
The bin sizes are consistent with those used in the binned QLFs.
The horizontal dotted lines indicate the completeness limits in each redshift bin, defined by \cite{Richards06a}.
\label{fig:nc}}
\end{figure}

\subsection{Selection Effect by Magnitude Cutoff\label{sec:seleffect}}

As described in Section \ref{sec:qlf}, our choice of the magnitude cutoff in optical ($i'<19.1$ mag) may result in the biased sample by excluding low-luminosity red quasars. 
Exclusion of the magnitude cutoff but retaining the other criteria in the main text yields an extended sample consisting of 108,198 quasars from the DR14Q catalog.
There are 14,549 red type-1 quasars, comprising 13.4\,\% of the total--slightly higher than the fraction reported in the main text.
This increase aligns with our hypothesis of a missing population of low-luminosity red quasars when a strict magnitude cutoff is applied.
Given the relatively shallower imaging depths in the WISE bands as discussed in Section \ref{sec:dr14q}, it is likely that this extended sample is constrained by the MIR magnitude cutoff, especially in the W4 band which has the shallowest imaging depth.

We explore the fraction of the missing population as a function of luminosity.
Figure \ref{fig:nc} shows the differential number counts of the red type-1 quasars in the $M_{i'}(z=2)$ magnitude after the $\ebv$ correction.
As expected, non-negligible fraction of red type-1 quasars are rejected around the completeness limit defined by \citeauthor{Richards06a} (\citeyear{Richards06a}; dotted line).
Specifically, in the two magnitude bins just brighter than this limit, fewer than 50\,\% of quasars are captured in each redshift bin.
We note that these selection efficiencies are not ideal for correcting the dereddened QLFs of red type-1 quasars in Section \ref{sec:qlf}, because the extended sample itself is also incomplete, being limited by MIR fluxes.

Meanwhile, our selection of type-1 quasars required detection in all WISE bands (Section \ref{sec:dr14q}), which may introduce a bias toward MIR-bright quasars.
Among the DR14Q, only 1943 quasars (4\,\% of 48,241) satisfies all other criteria but are not detected in at least one of the WISE bands.
Therefore, missing such quasars from our sample does not significantly affect our analysis given in this paper.

\begin{figure}
\centering
\epsscale{1.2}
\plotone{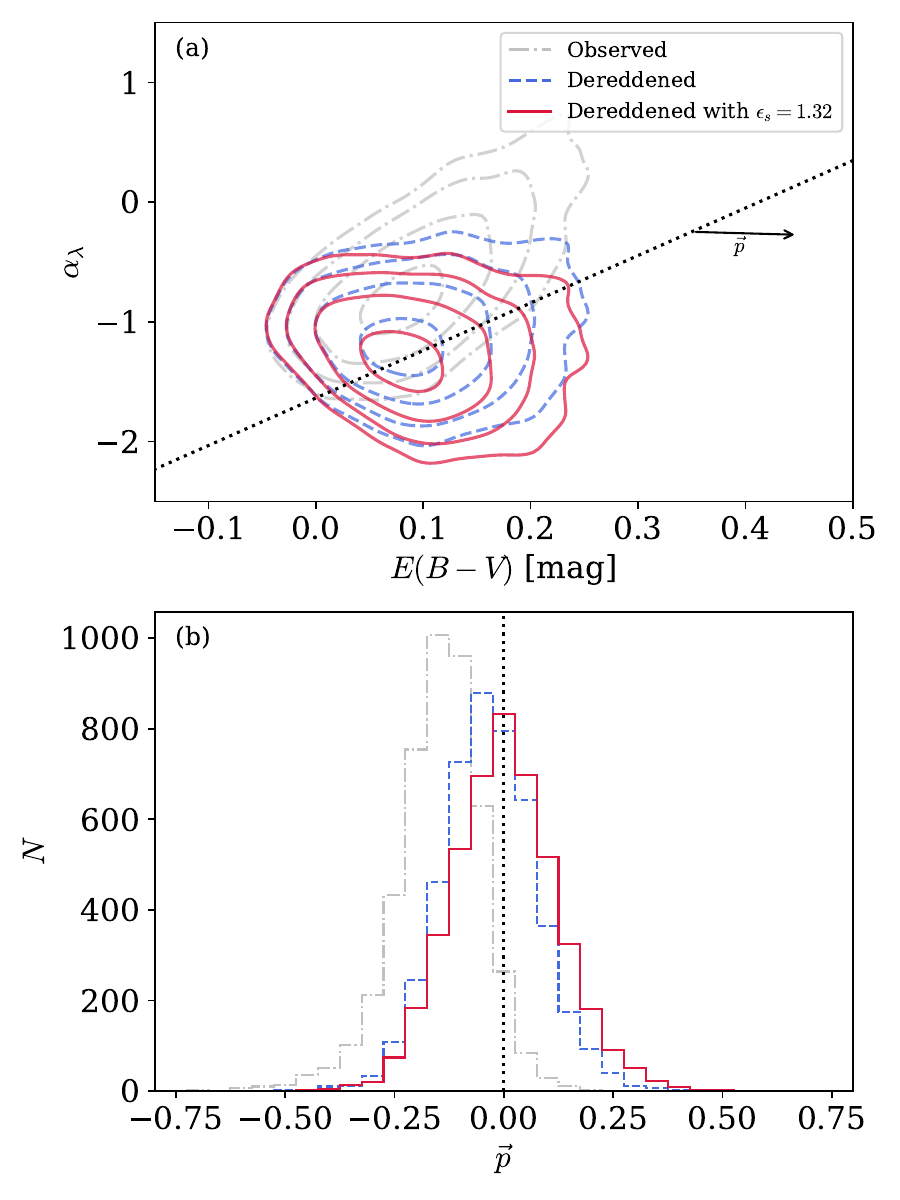}
\caption{
(a) $\ebv$-$\alplam$ distributions of red type-1 quasars.
The gray dash-dotted and blue dashed contours are the distributions for the observed and dereddened distributions, respectively.
The red contours indicates the distribution after the correction by $\ebv$ scaled with $\epsilon_s=1.32$.
The dotted line indicates the pseudo-correlation, and the arrow represents a vector $\vec{p}$, orthogonal to the pseudo-correlation.
(b) Projected $\ebv$-$\alplam$ distributions of red type-1 quasars onto a plane orthogonal to the pseudo-correlation.
The dotted vertical line marks the axis of the pseudo-correlation.
\label{fig:ebvscale}}
\end{figure}

\subsection{Model Bias in $\ebv$ Estimation\label{sec:ebvover}}

While we directly used $\ebv$ values obtained from our SED fitting for the $\ebv$ correction broadly used in this paper, the pseudo-correlation resulting from model limitations can lead to the over/underprediction of $\ebv$ and subsequent over/undercorrection.
Since the UV slope varies significantly among quasars even if they are unobscured, precise measurement of intrinsic $\ebv$ values is challenging.
We note that estimating $\ebv$ in different ways also poses challenges.
For example, $\ebv$ derived from line ratios and $\ebv$ obtained through the continuum slope do not show perfect agreement, exhibiting a scatter of approximately 0.2 \citep{Kim18}.

In Section \ref{sec:pseudo}, we derived a pseudo-correlation originated from our choice of SED model, serving as a reference point.
If the $\ebv$ values of the red type-1 quasars were measured precisely, the $\ebv$-corrected distribution of red type-1 quasars should be in line with the pseudo-correlation, as shown in the panel (c) of Figure \ref{fig:ebvslope}.
To address potential over/underestimation precisely, we introduced a scaling factor $\epsilon_s$ for $\ebv$; $\epsilon_s \times \ebv$.
This factor was optimized to ensure a good match between the distributions of red type-1 quasars and the pseudo-correlation.
In the panel (a) of Figure \ref{fig:ebvscale}, we present the $\ebv$-$\alplam$ distributions of red type-1 quasars.
The gray dash-dotted contours represent the observed distributions, while the blue dashed contours are the distributions after the $\ebv$ correction.
Note that these contours correspond to the red points in the panels (b) and (c) of Figure \ref{fig:ebvslope}, respectively.
We define $\vec{p}$ as the vector orthogonal to the pseudo-correlation (black arrow), oriented such that positive values correspond to increasing $\ebv$ and decreasing $\alplam$.
In the panel (b), we show the projected $\ebv$-$\alplam$ distribution for red type-1 quasars (gray dash-dotted histogram) onto a plane orthogonal to the pseudo-correlation.
The $\ebv$ correction shifts this distribution towards being centered around $\vec{p}\sim0$ (blue dashed histogram); however, the median value remains slightly negative at $-0.03$.
By applying a scaling factor of $\epsilon_s=1.32$, we align the median of the projected distribution with the pseudo-correlation (red histogram).
This $\epsilon_s$ value is marginally above 1, suggesting that our $\ebv$ measurements in the main text may have been slightly underestimated.
We further note that the median value for the blue quasars is $-0.01$, implying a possibility for a sligthly lower $\epsilon_s$ around 1.2.

If $\epsilon_s=1.32$ indeed contributes to $\ebv$ correction, the $M_{i'}(z=2)$ values of the red type-1 quasars slightly decrease (or become brighter), with a median decrease of 0.10 mag and a standard deviation of 0.14 mag.
However, this change is minor relative to the given QLF magnitude bin size of 0.3 mag.
Similarly, the $\mbh$ and $\redd$ estimations are also less affected by the potential under-dereddening.

\subsection{Inclusion of NIR Data}

Near-infrared (NIR) data have the potential to enhance the SED fitting process, particularly for capturing the hot dust components of low-redshift quasars and the host galaxy contribution of high-redshift quasars \citep{Boquien19,Kim23}.
However, we did not include NIR data in the main text to maximize the sky coverage and include the DR14Q sample as much as possible.
In this section, we investigate the effect of including NIR data on our main results.

Apart from optical and MIR photometry, \cite{Paris18} provides NIR photometry for DR14Q from two surveys: Two Micron All Sky Survey (2MASS; \citealt{Cutri03}) and UKIRT Infrared Deep Sky Survey (UKIDSS; \citealt{Lawrence07}).
2MASS photometry is obtained by matching DR14Q with sources in the 2MASS point source catalog, covering the entire sky but with relatively shallow imaging depths ($K<16.2$ mag) compared to SDSS and WISE photometry.
While 2MASS covers all sky, its imaging depths are quite shallow ($K<16.2$ mag) compared to the SDSS and WISE photometry we used.
On the other hand, UKIDSS photometry is derived from forced photometry on images, covering 3792 deg$^{2}$ of the sky with deeper imaging depths ($K<20.1$ mag).

We repeat the process in the main text, including the NIR data.
In addition to the selection criteria in Section \ref{sec:dr14q}, we set detection criteria of S/N$>3$ for NIR photometry.
Note that we utilize $JHK$-bands for 2MASS and $YJHK$-bands for UKIDSS.
The numbers of sample finalized after the SED fitting are listed in Table \ref{tbl:nir}.

\begin{deluxetable}{lccc}
\tabletypesize{\scriptsize}
\tablecaption{Effect of including NIR data\label{tbl:nir}}
\tablehead{
\colhead{} & \colhead{SDSS+WISE} & \colhead{+2MASS} & \colhead{+UKIDSS} 
}
\startdata
Bands & $u^{*}g'r'i'z'$+W1-W4 & +$JHK$ & +$YJHK$ \\
\hline
Sample  & 48,241 & 4875 & 10,482 \\
Red type-1 & 4610 (9.6\,\%) & 764 (16\,\%) & 1004 (9.6\,\%)\\
\hline
\hline
All & & & \\
$\Delta \ebv$ & ... & $0.000\pm0.002$ & $0.002\pm0.006$ \\
\hline
Red type-1 & & & \\
$\Delta \ebv$ & ... & $0.000\pm0.001$ & $-0.002\pm0.004$ \\
$\Delta \delta M_{i'}$ & ... & $0.000\pm0.006$ & $0.016\pm0.027$ \\
$\Delta \delta L_{3000}$ & ... & $0.000\pm0.002$ & $-0.006\pm0.011$ \\
$\Delta R_{\rm Q}(1.5\,\mu{\rm m})$ & ... & $0.000\pm0.027$ & $0.087\pm0.087$ \\
\enddata
\tablecomments{For the changes in the key parameters, the median and median absolute deviation values of the distributions in Figure \ref{fig:comp2muk} are listed.}
\end{deluxetable}

\begin{figure}
\centering
\epsscale{1.2}
\plotone{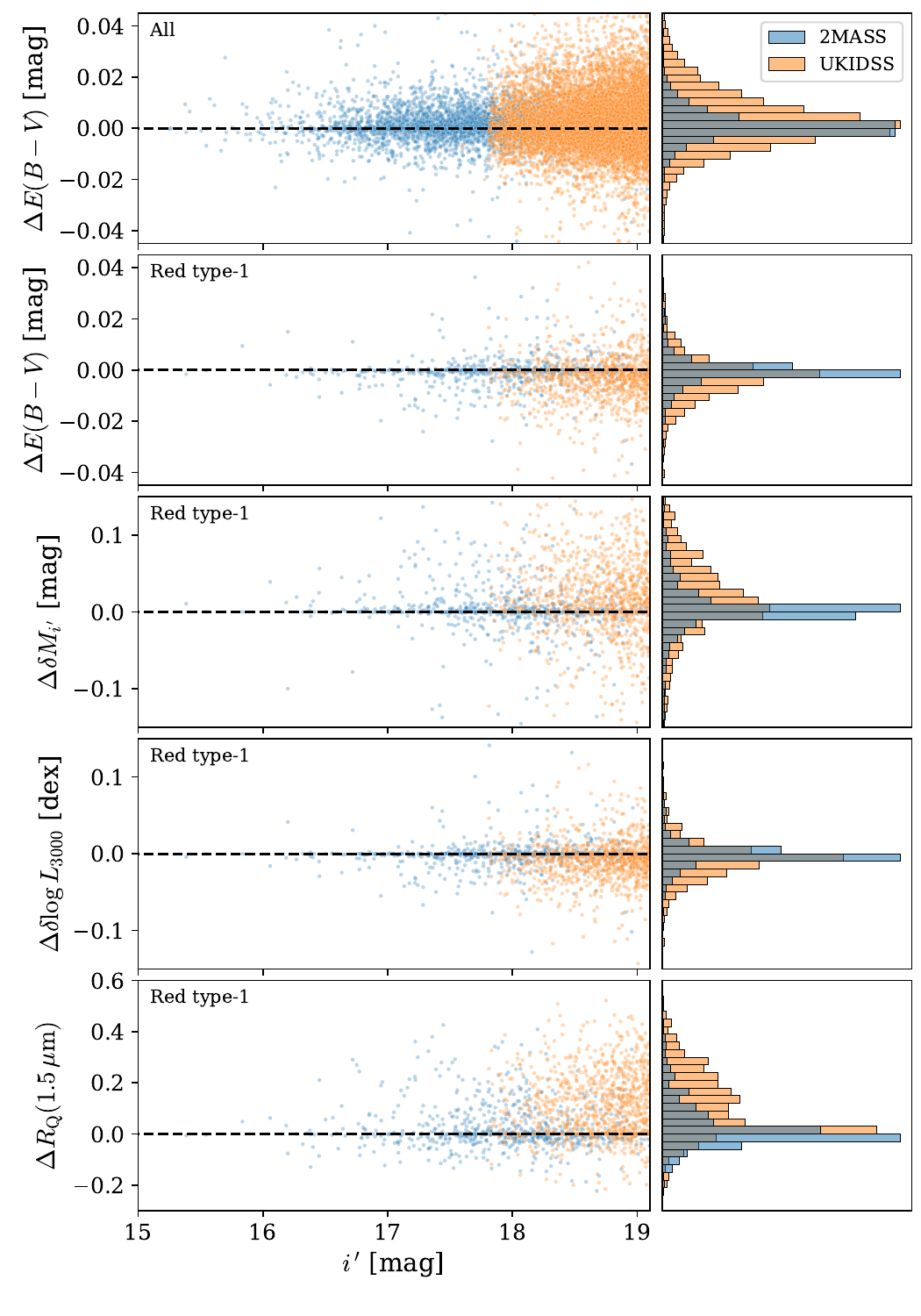}
\caption{
Change in key parameters depending on whether 2MASS (blue) or UKIDSS (orange) data are included, compared to the results in the main text.
The Top panel shows the change in $\ebv$ for our main sample matched to the samples with NIR data.
The other panels are only for the red type-1 quasars among the matched samples, shown in the order of changes in $\ebv$, $\delta M_{i'}$, $\delta \log L_{3000}$, and $R_{\rm Q}(1.5\,\mu{\rm m})$ from top to bottom.
\label{fig:comp2muk}}
\end{figure}

We first check the red type-1 quasar fractions of the cases including NIR data.
The fraction from the UKIDSS-included data is 9.6\,\%, similar to that from the optical+MIR data used in the main text.
On the other hand, the higher fraction of 2MASS-included data (16\,\%) is due to the selection effect with shallow imaging depths of the 2MASS survey.
If we apply $i'<17.2$ mag cut corresponding to the 2MASS detection limit ($K<16.2$ mag) based on the fiducial quasar SED template \citep{Richards06b}, the fraction decreases to 8.5\,\% (80/937) comparable to that of main sample.

Then we check the change in the key parameters related to our main results in Section \ref{sec:impact}, which are defined as follows.
$\Delta \ebv$ is the difference between the $\ebv$ values from the SED fitting with and without including the NIR data.
$\delta M_{i'}$ and $\delta \log L_{3000}$ are the degrees to which the $M_{i'}(z=2)$ and $\log L_{3000}$ values change according to the dereddening effect, respectively.
Similar to $\Delta \ebv$, the change in those values with and without including the NIR data are defined as $\Delta \delta M_{i'}$ and $\Delta \delta \log L_{3000}$.
We note that all these notations are defined as the result containing the NIR data minus the main text result.
Figure \ref{fig:comp2muk} shows the change in the key parameters for the matched samples, and the median and median absolute deviation values of the distributions are listed in Table \ref{tbl:nir}.

The $\Delta \ebv$ distributions for our main sample matched to the sample with NIR data exhibit very small scatters of $<0.01$ mag for both 2MASS and UKIDSS.
The scatters become slightly smaller if we consider only the red type-1 quasars that are considered for the dereddening effect according to the $\ebv$ values.
As a result, the changes in $\delta M_{i'}$ and $\delta \log L_{3000}$ are also minor, indicating that the lack of NIR data does not significantly impact our main results on the QLF and $\mbh$ estimations.
There is a little bias toward slightly lower $\ebv$ when including UKIDSS data, and resultant biases in dereddening effect.
However, there are only five outliers with $\Delta \delta M_{i'}>0.3$ mag, exceeding the magnitude bin size of 0.3 mag in QLF.
Moreover, since they move toward fainter bins where the number is high, there is no significant impact statistically. 
Similarly, the $\mbh$ estimations are also less affected by including the UKIDSS data with small changes in $\delta \log L_{3000}$ less than 0.1 dex.

The inclusion of NIR data improves determining the host galaxy fractions.
While the parameters related to the dereddening effect change minor, the fraction of quasar light is affected by including the NIR data, especially at the NIR wavelengths in the rest frame.
The bottom panel of Figure \ref{fig:comp2muk} shows the increasing trend in $\Delta R_{\rm Q}(1.5\,\mu{\rm m})$ as apparent magnitude goes fainter.
This means the underestimation of quasar light at this wavelength in our main text, implying NIR data is essential to measure pure quasar light in NIR and corresponding $\lbol$ as in \cite{Kim23}. 
Such a discrepancy becomes smaller at shorter wavelengths where quasar lights are dominant, and we note that this has less impact on our main results, as shown above.

\section{Summary}

In this study, we explored the presence of red type-1 quasars within a specific redshift range of $0.68 \leq z < 2.20$ in the SDSS DR14Q sample. By conducting SED fitting to their SDSS optical and AllWISE MIR photometric data, we identified approximately 10\,\% of type-1 quasars exhibiting red colors in the rest frame ($\Delta (g'-i') > 0.3$ mag), indicative of potential dust obscuration effects.
We observed that these red type-1 quasars are inherently biased towards higher MIR luminosities, with their fraction peaking at around 40\,\% at the highest luminosity, similar to findings from MIR-based surveys. 
Interestingly, this prevalence remains relatively stable across different redshifts.

To assess the impact of dust obscuration on quasar properties, we derived dereddened luminosities from the measured $\ebv$ values obtained from SED fitting. 
Our analysis revealed a subtle increase in the number density of quasars at the brighter end of the QLFs, although the differences were not significant. 
Similarly, the changes in $\mbh$ estimation appeared to be negligible.
It's important to note that our study relies on DR14Q, primarily detecting quasars in optical wavelengths, potentially missing populations with $\ebv$ values larger than 0.6. Future surveys, particularly those targeting MIR wavelengths like SPHEREx, may uncover more dust-obscured quasars with broad emission lines, which could significantly alter our understanding of fundamental quasar properties, including the QLF and $\mbh$ estimation.

\acknowledgments

We thank the anonymous referee for the useful comments that helped us improve the draft.
Y. K. was supported by the National Research Foundation of Korea (NRF) grant funded by the Korean government (MSIT) (No. 2021R1C1C2091550).
D. K. was supported by the National Research Foundation of Korea (NRF) grant funded by the Korea government (MSIT) (No. 2021R1C1C1013580 and 2022R1A4A3031306).
M. I. acknowledges the support from the National Research Foundation of Korea (NRF) grants, No. 2020R1A2C3011091, and No. 2021M3F7A1084525, funded by the Korea government (MSIT).
M. K. acknowledges the support from the National Research Foundation of Korea (NRF) grants (No. RS-2024-00347548).

This publication makes use of data products from the Wide-field Infrared Survey Explorer, which is a joint project of the University of California, Los Angeles, and the Jet Propulsion Laboratory/California Institute of Technology, funded by the National Aeronautics and Space Administration.

This publication makes use of data products from the Two Micron All Sky Survey, which is a joint project of the University of Massachusetts and the Infrared Processing and Analysis Center/California Institute of Technology, funded by the National Aeronautics and Space Administration and the National Science Foundation.

\software{MPFIT \citep{Markwardt09}; Astropy \citep{Astropy22}}




\bigskip

\begin{thebibliography}{}

\bibitem[Aird et al.(2015)]{Aird15} Aird, J., Coil, A.~L., Georgakakis, A., et al.\ 2015, \mnras, 451, 1892. doi:10.1093/mnras/stv1062

\bibitem[Assef et al.(2010)]{Assef10} Assef, R.~J., Kochanek, C.~S., Brodwin, M., et al.\ 2010, \apj, 713, 970. doi:10.1088/0004-637X/713/2/970

\bibitem[Assef et al.(2015)]{Assef15} Assef, R.~J., Eisenhardt, P.~R.~M., Stern, D., et al.\ 2015, \apj, 804, 27. doi:10.1088/0004-637X/804/1/27

\bibitem[Astropy Collaboration et al.(2022)]{Astropy22} Astropy Collaboration, Price-Whelan, A.~M., Lim, P.~L., et al.\ 2022, \apj, 935, 167. doi:10.3847/1538-4357/ac7c74

\bibitem[Avni \& Bahcall(1980)]{Avni80} Avni, Y. \& Bahcall, J.~N.\ 1980, \apj, 235, 694. doi:10.1086/157673

\bibitem[Baldwin et al.(1981)]{Baldwin81} Baldwin, J.~A., Phillips, M.~M., \& Terlevich, R.\ 1981, \pasp, 93, 5. doi:10.1086/130766

\bibitem[Banerji et al.(2015)]{Banerji15} Banerji, M., Alaghband-Zadeh, S., Hewett, P.~C., et al.\ 2015, \mnras, 447, 3368. doi:10.1093/mnras/stu2649

\bibitem[Bentz et al.(2009)]{Bentz09} Bentz, M.~C., Walsh, J.~L., Barth, A.~J., et al.\ 2009, \apj, 705, 199. doi:10.1088/0004-637X/705/1/199

\bibitem[Bentz et al.(2013)]{Bentz13} Bentz, M.~C., Denney, K.~D., Grier, C.~J., et al.\ 2013, \apj, 767, 149. doi:10.1088/0004-637X/767/2/149

\bibitem[Blanton et al.(2001)]{Blanton01} Blanton, M.~R., Dalcanton, J., Eisenstein, D., et al.\ 2001, \aj, 121, 2358. doi:10.1086/320405

\bibitem[Boquien et al.(2019)]{Boquien19} Boquien, M., Burgarella, D., Roehlly, Y., et al.\ 2019, \aap, 622, A103. doi:10.1051/0004-6361/201834156

\bibitem[Boyle et al.(2000)]{Boyle00} Boyle, B.~J., Shanks, T., Croom, S.~M., et al.\ 2000, \mnras, 317, 1014. doi:10.1046/j.1365-8711.2000.03730.x

\bibitem[Calzetti et al.(2000)]{Calzetti00} Calzetti, D., Armus, L., Bohlin, R.~C., et al.\ 2000, \apj, 533, 682. doi:10.1086/308692

\bibitem[Conroy \& White(2013)]{Conroy13} Conroy, C. \& White, M.\ 2013, \apj, 762, 70. doi:10.1088/0004-637X/762/2/70

\bibitem[Croom et al.(2004)]{Croom04} Croom, S.~M., Smith, R.~J., Boyle, B.~J., et al.\ 2004, \mnras, 349, 1397. doi:10.1111/j.1365-2966.2004.07619.x


\bibitem[Croom et al.(2009)]{Croom09} Croom, S.~M., Richards, G.~T., Shanks, T., et al.\ 2009b, \mnras, 399, 1755. doi:10.1111/j.1365-2966.2009.15398.x 


\bibitem[Cutri et al.(2003)]{Cutri03} Cutri, R.~M., Skrutskie, M.~F., van Dyk, S., et al.\ 2003, VizieR Online Data Catalog, II/246

\bibitem[Cutri et al.(2021)]{Cutri21} Cutri, R.~M., Wright, E.~L., Conrow, T., et al.\ 2021, VizieR Online Data Catalog, II/328

\bibitem[Dietrich et al.(2002)]{Dietrich02} Dietrich, M., Hamann, F., Shields, J.~C., et al.\ 2002, \apj, 581, 912 

\bibitem[Dor{\'e} et al.(2014)]{Dore14} Dor{\'e}, O., Bock, J., Ashby, M., et al.\ 2014, arXiv:1412.4872. doi:10.48550/arXiv.1412.4872

\bibitem[Glikman et al.(2004)]{Glikman04} Glikman, E., Gregg, M.~D., Lacy, M., et al.\ 2004, \apj, 607, 60. doi:10.1086/383305

\bibitem[Glikman et al.(2012)]{Glikman12} Glikman, E., Urrutia, T., Lacy, M., et al.\ 2012, \apj, 757, 51. doi:10.1088/0004-637X/757/1/51

\bibitem[Glikman et al.(2015)]{Glikman15} Glikman, E., Simmons, B., Mailly, M., et al.\ 2015, \apj, 806, 218. doi:10.1088/0004-637X/806/2/218

\bibitem[Glikman et al.(2018)]{Glikman18} Glikman, E., Lacy, M., LaMassa, S., et al.\ 2018, \apj, 861, 37. doi:10.3847/1538-4357/aac5d8

\bibitem[Gunn et al.(2006)]{Gunn06} Gunn, J.~E., Siegmund, W.~A., Mannery, E.~J., et al.\ 2006, \aj, 131, 2332. doi:10.1086/500975

\bibitem[Guo et al.(2018)]{Guo18} Guo, H., Shen, Y., \& Wang, S.\ 2018, Astrophysics Source Code Library. ascl:1809.008

\bibitem[Hickox \& Alexander(2018)]{Hickox18} Hickox, R.~C. \& Alexander, D.~M.\ 2018, \araa, 56, 625. doi:10.1146/annurev-astro-081817-051803

\bibitem[Ho et al.(2012)]{Ho12} Ho, L.~C., Goldoni, P., Dong, X.-B., et al.\ 2012, \apj, 754, 11. doi:10.1088/0004-637X/754/1/11

\bibitem[Hopkins et al.(2008)]{Hopkins08} Hopkins, P.~F., Hernquist, L., Cox, T.~J., et al.\ 2008, \apjs, 175, 356. doi:10.1086/524362

\bibitem[Inoue et al.(2014)]{Inoue14} Inoue, A.~K., Shimizu, I., Iwata, I., et al.\ 2014, \mnras, 442, 1805. doi:10.1093/mnras/stu936

\bibitem[Jiang et al.(2008)]{Jiang08} Jiang, L., Fan, X., Annis, J., et al.\ 2008, \aj, 135, 1057. doi:10.1088/0004-6256/135/3/1057

\bibitem[Jiang et al.(2009)]{Jiang09} Jiang, L., Fan, X., Bian, F., et al.\ 2009, \aj, 138, 305. doi:10.1088/0004-6256/138/1/305

\bibitem[Jiang et al.(2016)]{Jiang16} Jiang, L., McGreer, I.~D., Fan, X., et al.\ 2016, \apj, 833, 222. doi:10.3847/1538-4357/833/2/222

\bibitem[Kelly \& Shen(2013)]{Kelly13} Kelly, B.~C. \& Shen, Y.\ 2013, \apj, 764, 45. doi:10.1088/0004-637X/764/1/45

\bibitem[Kim et al.(2015a)]{Kim15} Kim, Y., Im, M., Jeon, Y., et al.\ 2015a, \apjl, 813, L35. doi:10.1088/2041-8205/813/2/L35

\bibitem[Kim et al.(2020)]{Kim20} Kim, Y., Im, M., Jeon, Y., et al.\ 2020, \apj, 904, 111. doi:10.3847/1538-4357/abc0ea

\bibitem[Kim et al.(2022)]{Kim22} Kim, Y., Im, M., Jeon, Y., et al.\ 2022, \aj, 164, 114. doi:10.3847/1538-3881/ac81c8

\bibitem[Kim \& Im(2021)]{KimIm21} Kim, Y. \& Im, M.\ 2021, \apjl, 910, L11. doi:10.3847/2041-8213/abed58

\bibitem[Kim et al.(2015b)]{KimD15} Kim, D., Im, M., Glikman, E., et al.\ 2015b, \apj, 812, 66. doi:10.1088/0004-637X/812/1/66

\bibitem[Kim \& Im(2018)]{KimIm18} Kim, D. \& Im, M.\ 2018, \aap, 610, A31. doi:10.1051/0004-6361/201731963

\bibitem[Kim et al.(2018)]{Kim18} Kim, D., Im, M., Canalizo, G., et al.\ 2018, \apjs, 238, 37. doi:10.3847/1538-4365/aadfd5

\bibitem[Kim et al.(2023)]{Kim23} Kim, D., Im, M., Kim, M., et al.\ 2023, \apj, 954, 156. doi:10.3847/1538-4357/aceb5e

\bibitem[Kim et al.(2024a)]{Kim24a} Kim, D., Kim, Y., Im, M., et al.\ 2024a, arXiv:2408.03228

\bibitem[Kim et al.(2024b)]{Kim24b} Kim, D., Im, M., Lim, G., et al.\ 2024b, Journal of Korean Astronomical Society, 57, 95. doi:10.5303/JKAS.2024.57.1.95

\bibitem[Krawczyk et al.(2013)]{Krawczyk13} Krawczyk, C.~M., Richards, G.~T., Mehta, S.~S., et al.\ 2013, \apjs, 206, 4. doi:10.1088/0067-0049/206/1/4

\bibitem[Kulkarni et al.(2019)]{Kulkarni19} Kulkarni, G., Worseck, G., \& Hennawi, J.~F.\ 2019, \mnras, 488, 1035. doi:10.1093/mnras/stz1493

\bibitem[Lacy et al.(2013)]{Lacy13} Lacy, M., Ridgway, S.~E., Gates, E.~L., et al.\ 2013, \apjs, 208, 24. doi:10.1088/0067-0049/208/2/24

\bibitem[Lacy et al.(2015)]{Lacy15} Lacy, M., Ridgway, S.~E., Sajina, A., et al.\ 2015, \apj, 802, 102. doi:10.1088/0004-637X/802/2/102

\bibitem[Lawrence et al.(2007)]{Lawrence07} Lawrence, A., Warren, S.~J., Almaini, O., et al.\ 2007, \mnras, 379, 1599 

\bibitem[Le et al.(2020)]{Le20} Le, H.~A.~N., Woo, J.-H., \& Xue, Y.\ 2020, \apj, 901, 35. doi:10.3847/1538-4357/abada0

\bibitem[Lusso et al.(2015)]{Lusso15} Lusso, E., Worseck, G., Hennawi, J.~F., et al.\ 2015, \mnras, 449, 4204. doi:10.1093/mnras/stv516

\bibitem[Lyke et al.(2020)]{Lyke20} Lyke, B.~W., Higley, A.~N., McLane, J.~N., et al.\ 2020, \apjs, 250, 8. doi:10.3847/1538-4365/aba623

\bibitem[Mainzer et al.(2011)]{Mainzer11} Mainzer, A., Bauer, J., Grav, T., et al.\ 2011, \apj, 731, 53. doi:10.1088/0004-637X/731/1/53

\bibitem[Markwardt(2009)]{Markwardt09} Markwardt, C.~B.\ 2009, Astronomical Data Analysis Software and Systems XVIII, 411, 251. doi:10.48550/arXiv.0902.2850

\bibitem[McGreer et al.(2013)]{McGreer13} McGreer, I.~D., Jiang, L., Fan, X., et al.\ 2013, \apj, 768, 105. doi:10.1088/0004-637X/768/2/105

\bibitem[Montero-Dorta \& Prada(2009)]{Montero09} Montero-Dorta, A.~D. \& Prada, F.\ 2009, \mnras, 399, 1106. doi:10.1111/j.1365-2966.2009.15197.x

\bibitem[Page \& Carrera(2000)]{Page00} Page, M.~J. \& Carrera, F.~J.\ 2000, \mnras, 311, 433. doi:10.1046/j.1365-8711.2000.03105.x

\bibitem[Palanque-Delabrouille et al.(2013)]{Palanque13} Palanque-Delabrouille, N., Magneville, C., Y{\`e}che, C., et al.\ 2013, \aap, 551, A29. doi:10.1051/0004-6361/201220379

\bibitem[Palanque-Delabrouille et al.(2016)]{Palanque16} Palanque-Delabrouille, N., Magneville, C., Y{\`e}che, C., et al.\ 2016, \aap, 587, A41. doi:10.1051/0004-6361/201527392

\bibitem[P{\^a}ris et al.(2012)]{Paris12} P{\^a}ris, I., Petitjean, P., Aubourg, {\'E}., et al.\ 2012, \aap, 548, A66. doi:10.1051/0004-6361/201220142

\bibitem[P{\^a}ris et al.(2014)]{Paris14} P{\^a}ris, I., Petitjean, P., Aubourg, {\'E}., et al.\ 2014, \aap, 563, A54. doi:10.1051/0004-6361/201322691

\bibitem[P{\^a}ris et al.(2017)]{Paris17} P{\^a}ris, I., Petitjean, P., Ross, N.~P., et al.\ 2017, \aap, 597, A79. doi:10.1051/0004-6361/201527999

\bibitem[P{\^a}ris et al.(2018)]{Paris18} P{\^a}ris, I., Petitjean, P., Aubourg, {\'E}., et al.\ 2018, \aap, 613, A51. doi:10.1051/0004-6361/201732445

\bibitem[Polletta et al.(2008)]{Polletta08} Polletta, M., Weedman, D., H{\"o}nig, S., et al.\ 2008, \apj, 675, 960. doi:10.1086/524343

\bibitem[Rakshit et al.(2020)]{Rakshit20} Rakshit, S., Stalin, C.~S., \& Kotilainen, J.\ 2020, \apjs, 249, 17. doi:10.3847/1538-4365/ab99c5

\bibitem[Ren et al.(2020)]{Ren20} Ren, K., Trenti, M., \& Di Matteo, T.\ 2020, \apj, 894, 124. doi:10.3847/1538-4357/ab86ab

\bibitem[Richards et al.(2003)]{Richards03} Richards, G.~T., Hall, P.~B., Vanden Berk, D.~E., et al.\ 2003, \aj, 126, 1131. doi:10.1086/377014

\bibitem[Richards et al.(2006a)]{Richards06a} Richards, G.~T., Strauss, M.~A., Fan, X., et al.\ 2006a, \aj, 131, 2766. doi:10.1086/503559

\bibitem[Richards et al.(2006b)]{Richards06b} Richards, G.~T., Lacy, M., Storrie-Lombardi, L.~J., et al.\ 2006b, \apjs, 166, 470. doi:10.1086/506525

\bibitem[Rose et al.(2013)]{Rose13} Rose, M., Tadhunter, C.~N., Holt, J., et al.\ 2013, \mnras, 432, 2150. doi:10.1093/mnras/stt564

\bibitem[Ross et al.(2013)]{Ross13} Ross, N.~P., McGreer, I.~D., White, M., et al.\ 2013, \apj, 773, 14. doi:10.1088/0004-637X/773/1/14

\bibitem[Schlafly \& Finkbeiner(2011)]{Schlafly11} Schlafly, E.~F. \& Finkbeiner, D.~P.\ 2011, \apj, 737, 103. doi:10.1088/0004-637X/737/2/103

\bibitem[Schneider et al.(2002)]{Schneider02} Schneider, D.~P., Richards, G.~T., Fan, X., et al.\ 2002, \aj, 123, 567. doi:10.1086/338434

\bibitem[Schneider et al.(2003)]{Schneider03} Schneider, D.~P., Fan, X., Hall, P.~B., et al.\ 2003, \aj, 126, 2579. doi:10.1086/379174

\bibitem[Schneider et al.(2005)]{Schneider05} Schneider, D.~P., Hall, P.~B., Richards, G.~T., et al.\ 2005, \aj, 130, 367. doi:10.1086/431156

\bibitem[Schneider et al.(2007)]{Schneider07} Schneider, D.~P., Hall, P.~B., Richards, G.~T., et al.\ 2007, \aj, 134, 102. doi:10.1086/518474

\bibitem[Schneider et al.(2010)]{Schneider10} Schneider, D.~P., Richards, G.~T., Hall, P.~B., et al.\ 2010, \aj, 139, 2360. doi:10.1088/0004-6256/139/6/2360

\bibitem[Selsing et al.(2016)]{Selsing16} Selsing, J., Fynbo, J.~P.~U., Christensen, L., et al.\ 2016, \aap, 585, A87. doi:10.1051/0004-6361/201527096

\bibitem[Shen et al.(2008)]{Shen08} Shen, Y., Greene, J.~E., Strauss, M.~A., et al.\ 2008, \apj, 680, 169. doi:10.1086/587475

\bibitem[Shen et al.(2011)]{Shen11} Shen, Y., Richards, G.~T., Strauss, M.~A., et al.\ 2011, \apjs, 194, 45. doi:10.1088/0067-0049/194/2/45

\bibitem[Shen \& Kelly(2012)]{Shen12} Shen, Y. \& Kelly, B.~C.\ 2012, \apj, 746, 169. doi:10.1088/0004-637X/746/2/169

\bibitem[Shen et al.(2013)]{Shen13} Shen, Y., Liu, X., Loeb, A., et al.\ 2013, \apj, 775, 49. doi:10.1088/0004-637X/775/1/49

\bibitem[Shen et al.(2019)]{Shen19} Shen, Y., Hall, P.~B., Horne, K., et al.\ 2019, \apjs, 241, 34. doi:10.3847/1538-4365/ab074f

\bibitem[Shen et al.(2020)]{Shen20} Shen, X., Hopkins, P.~F., Faucher-Gigu{\`e}re, C.-A., et al.\ 2020, \mnras, 495, 3252. doi:10.1093/mnras/staa1381

\bibitem[Shen et al.(2023)]{Shen23} Shen, Y., Grier, C.~J., Horne, K., et al.\ 2023, arXiv:2305.01014. doi:10.48550/arXiv.2305.01014

\bibitem[Telfer et al.(2002)]{Telfer02} Telfer, R.~C., Zheng, W., Kriss, G.~A., et al.\ 2002, \apj, 565, 773. doi:10.1086/324689

\bibitem[Ueda et al.(2014)]{Ueda14} Ueda, Y., Akiyama, M., Hasinger, G., et al.\ 2014, \apj, 786, 104. doi:10.1088/0004-637X/786/2/104

\bibitem[Urrutia et al.(2008)]{Urrutia08} Urrutia, T., Lacy, M., \& Becker, R.~H.\ 2008, \apj, 674, 80. doi:10.1086/523959

\bibitem[Urrutia et al.(2009)]{Urrutia09} Urrutia, T., Becker, R.~H., White, R.~L., et al.\ 2009, \apj, 698, 1095. doi:10.1088/0004-637X/698/2/1095

\bibitem[Urrutia et al.(2012)]{Urrutia12} Urrutia, T., Lacy, M., Spoon, H., et al.\ 2012, \apj, 757, 125. doi:10.1088/0004-637X/757/2/125

\bibitem[Vanden Berk et al.(2001)]{Vanden01} Vanden Berk, D.~E., Richards, G.~T., Bauer, A., et al.\ 2001, \aj, 122, 549 

\bibitem[Vanden Berk et al.(2005)]{Vanden05} Vanden Berk, D.~E., Schneider, D.~P., Richards, G.~T., et al.\ 2005, \aj, 129, 2047. doi:10.1086/427856

\bibitem[Vestergaard \& Wilkes(2001)]{Vestergaard01} Vestergaard, M. \& Wilkes, B.~J.\ 2001, \apjs, 134, 1. doi:10.1086/320357

\bibitem[Vestergaard \& Peterson(2006)]{Vestergaard06} Vestergaard, M. \& Peterson, B.~M.\ 2006, \apj, 641, 689. doi:10.1086/500572

\bibitem[Vestergaard \& Osmer(2009)]{VO09} Vestergaard, M. \& Osmer, P.~S.\ 2009, \apj, 699, 800. doi:10.1088/0004-637X/699/1/800

\bibitem[Vijarnwannaluk et al.(2022)]{Vijarnwannaluk22} Vijarnwannaluk, B., Akiyama, M., Schramm, M., et al.\ 2022, \apj, 941, 97. doi:10.3847/1538-4357/ac9c07

\bibitem[Wright et al.(2010)]{Wright10} Wright, E.~L., Eisenhardt, P.~R.~M., Mainzer, A.~K., et al.\ 2010, \aj, 140, 1868. doi:10.1088/0004-6256/140/6/1868

\bibitem[York et al.(2000)]{York00} York, D.~G., Adelman, J., Anderson, J.~E., et al.\ 2000, \aj, 120, 1579. doi:10.1086/301513



\end{thebibliography}
\end{document}